\documentclass[a4paper,12pt]{article}
\pdfoutput=1

\usepackage{jheppub,fancyhdr,graphicx,epstopdf,color,amsmath,cases,slashed,hyperref}

\newcommand{\beq}{\begin{eqnarray}}% can be used as {equation} or {eqnarray}
\newcommand{\eeq}{\end{eqnarray}}

\def\ltap{\ \raise.3ex\hbox{$<$\kern-.75em\lower1ex\hbox{$\sim$}}\ }
\def\gtap{\ \raise.3ex\hbox{$>$\kern-.75em\lower1ex\hbox{$\sim$}}\ }

\def\be{\begin{equation}}
\def\ee{\end{equation}}
\def\bea{\begin{eqnarray}}
\def\eea{\end{eqnarray}}

\definecolor{red1}{cmyk}{0,1,1,0.3}

\title{\boldmath Cornering pseudoscalar-mediated dark matter with the LHC and cosmology}

\author[a]{Shankha~Banerjee}
\author[b]{\!\!, Daniele~Barducci}
\author[a]{\!\!, Genevi\`eve~B\'elanger}
\author[c,d]{\!\!, Benjamin~Fuks}
\author[c]{\!\!\!, Andreas~Goudelis}
\author[a]{\! and Bryan~Zaldivar}

\affiliation[a]{LAPTh, Universit\'e Savoie Mont Blanc, CNRS, BP~110, F-74941 Annecy-le-Vieux, France}
\affiliation[b]{SISSA and INFN, Sezione di Trieste, via Bonomea 265, 34136 Trieste, Italy}
\affiliation[c]{LPTHE, Sorbonne Universit\'es, UPMC, UMR 7589 - CNRS, F-75252 Paris Cedex, France}
\affiliation[d]{Institut Universitaire de France, 103 boulevard Saint-Michel, 75005 Paris, France}

\emailAdd{banerjee@lapth.cnrs.fr}
\emailAdd{daniele.barducci@sissa.it}
\emailAdd{belanger@lapth.cnrs.fr}
\emailAdd{fuks@lpthe.jussieu.fr}
\emailAdd{andreas.goudelis@lpthe.jussieu.fr}
\emailAdd{zaldivar@lapth.cnrs.fr}

\abstract
{
Models in which dark matter particles communicate with the visible sector through a pseudoscalar
mediator are well-motivated both from a theoretical and from a phenomenological standpoint. With direct detection
bounds being typically subleading in such scenarios, the main constraints stem either from collider searches for
dark matter, or from indirect detection experiments. 
However, LHC searches for the mediator particles themselves can not only compete with
-- or even supersede -- the reach of direct collider dark matter probes, but they can also
test scenarios in which traditional monojet searches become irrelevant, especially
when the mediator cannot  decay on-shell into dark matter particles or its decay is suppressed. In this work we perform a detailed analysis of a
pseudoscalar-mediated dark matter simplified model, taking into account a large
set of collider constraints and concentrating on the parameter space regions
favoured by cosmological and astrophysical data. We find that mediator
masses above 100-200~GeV are essentially excluded by LHC searches in the case of
large couplings to the top quark, while forthcoming collider and astrophysical measurements will further constrain
the available parameter space.
}

\begin{document}

\date\today

\begin{flushright}
\hspace{3cm} 
LAPTH-014/17\\
SISSA-22-2017-FISI
\end{flushright}

\maketitle
\flushbottom

%%%%%%%%%%%%%%%%%%%%%%%%%%%%%%%%%%%%%%%%%%%%%%%%%%%%%%%%%%%%%%%%%%%%

\section{Introduction}\label{sec:introduction}

Searches for dark matter (DM) particles constitute one of the main physics objectives
of the LHC, their prime signature being associated with the presence of missing
transverse energy ($\slashed{E}_T$) in the collision final state. Traditionally
conducted within specific dark matter models like supersymmetry
that can give rise to a variety of final states involving $\slashed{E}_T$, LHC
analyses have lately shifted towards more model-independent
approaches~\cite{Abercrombie:2015wmb}. One such approach is based on effective
field theories (EFT)~\cite{Goodman:2010yf,Goodman:2010ku} which, despite its simplicity, can draw a misleading picture at LHC energies when TeV-scale new degrees of freedom cannot be integrated out~\cite{Busoni:2013lha,Busoni:2014sya,Busoni:2014haa}. Another approach relies on so-called \emph{simplified models}, \textit{i.e.} simple frameworks which
extend the Standard Model (SM) by two particles, a dark matter candidate as well as a
state that mediates the dark matter interactions with the visible 
sector~\cite{Abdallah:2015ter,Buckley:2014fba,Harris:2014hga}. Such frameworks palliate some of the drawbacks of the EFT approach, at the price of introducing only a handful of additional free parameters. Moreover, simplified models capture, with a minimal set of assumptions, some important features of more ultraviolet-complete (UV) theories and, perhaps more importantly, they can provide a (semi-)consistent framework in order to analyse the
experimental results~\cite{Buchmueller:2013dya}.
These two approaches are indeed complementary, since they explore different regions of the mediating particle's mass scale~\footnote{In some cases the EFT approach is the only suitable description. This happens for example in the presence of strongly coupled UV completions~\cite{Bruggisser:2016nzw}.} and consistently set EFT limits can in principle be reinterpreted in any specific underlying model~\cite{Racco:2015dxa}. 
Independently of the theoretical framework, the most well studied
mo\-del-in\-de\-pen\-dent dark matter signatures at the LHC have been the mono-X
ones~\cite{Feng:2005gj,Birkedal:2004xn,Bai:2010hh,Fox:2011fx,Bell:2012rg,Petrov:2013nia,Bai:2012xg},
in which a pair of dark matter particles is produced in association with a
single energetic visible object. Amongst all mono-X searches, the monojet one has garnered
the most attention given the relative
magnitude of the strong coupling with respect to the electroweak one. This
channel was shown to provide powerful constraints on dark matter models,
especially in cases where direct detection experiments become inefficient~\cite{Aaboud:2016tnv,Sirunyan:2017hci}.
Barring somehow singular kinematic configurations that can occur in freeze-out
scenarios~\cite{Griest:1990kh}, the two most notable situations in which direct detection constraints fall short concern models in which the dark matter particles can be lighter than a few GeV, where direct detection searches suffer from recoil energy threshold limitations \cite{Akerib:2016vxi}, or models in which the spin-independent dark matter-nucleon scattering cross section is suppressed due to the Lorentz structure of the underlying theory.
This suppression can occur in models of axial-vector mediated fermionic
dark matter or, which is the topic of this work, in scenarios featuring a pseudoscalar-mediated fermionic dark matter candidate in which the dark matter-nucleon scattering cross section is suppressed by the momentum transfer involved in the reaction.

If the pseudoscalar mediator, moreover, couples to the Standard Model fermions
proportionally to their mass (an assumption motivated both by the success of
Minimal Flavour Violation~\cite{DAmbrosio:2002vsn,Buras:2000dm} and by ultraviolet considerations), it should be
mostly produced through gluon fusion, analogously to Standard Model Higgs boson
production. Gluon fusion processes typically
exhibit an enriched jet activity compared to quark-antiquark annihilation ones. As a
consequence, searches for final state signatures comprised of a multijet system
accompanied by $\slashed{E}_T$ could potentially lead to more
stringent constraints than those originating from pure monojet searches, as
first argued in Ref.~\cite{Buchmueller:2015eea} on the basis of 8~TeV LHC data.
For this reason, more recent monojet LHC analyses at 13~TeV allow for
an additional hadronic activity in their selection. In this work we investigate the constraints
on the parameter space of a simplified  pseudoscalar-mediated fermionic
dark matter model stemming from both monojet and multijet plus $\slashed{E}_T$
searches at the LHC. We quantify the impact of
higher-order QCD corrections and we further study the bounds originating from complementary LHC dark matter searches such as the
associated production of an invisibly-decaying pseudoscalar mediator with top or
bottom quark pairs, also highlighting the future prospects for these searches.

Besides collider searches for dark matter in channels with large
$\slashed{E}_T$, resonance searches by means of signatures made up solely
of visible objects could be useful to constrain the properties of the mediator
connecting the dark and visible sectors~\cite{Chala:2015ama,Arina:2016cqj,Kraml:2017atm}. We therefore
revisit LHC studies probing a potential new physics resonance
decaying into a pair of tau leptons~\cite{Aaboud:2016cre,CMS-PAS-HIG-16-037},
photons~\cite{ATLAS:2016eeo} or top quarks~\cite{Aad:2015fna}. 
Moreover, as we investigate scenarios with a large coupling of the mediator
to top quarks, we also assess the sensitivity of the direct measurement of the
top quark pair production cross section to the new physics parameter space. This observable turns out
to play a key role even when a top-antitop pair is produced via an off-shell
pseudoscalar exchange.

As the possible observation of enhanced missing energy signals (or, even more
so, of a new resonance) does not guarantee their cosmological relevance, we
compare the LHC constraints on the parameter space with the corresponding
regions that are phenomenologically viable from a cosmological and astrophysical standpoint. Concretely, we 
investigate bounds arising from the dark matter relic density
as well as from indirect searches such as the Fermi-LAT searches for dark matter-induced gamma-rays in Dwarf Spheroidal Galaxies~\cite{Ackermann:2015zua} and for spectral features at the Galactic Centre \cite{Ackermann:2015lka}, but also the AMS-02 searches for antiprotons \cite{Aguilar:2016kjl}. Future prospects for indirect detection experiments are also discussed.

The paper is organised as follows: In section~\ref{sec:model} we present the
theoretical framework for our study. In section~\ref{sec:lhc} we describe the
various collider constraints that we consider, whereas
section~\ref{sec:darkmatter} is dedicated to dark matter astrophysical observables. Our results are presented in section~\ref{sec:resultsanddiscussion}
while in section~\ref{sec:outlook} we provide our conclusions. Two appendices follow,
discussing some technical aspects of our analysis.

\section{Model description}\label{sec:model}

We consider a new physics scenario in which the Standard Model field content is
extended by a Majorana fermionic field $\chi$ of mass $m_\chi$, which plays the
role of a dark matter candidate, and a pseudoscalar field $A$ of mass $m_A$,
which mediates the interactions
of $\chi$ with the Standard Model. Both particles are taken to be singlets under
the Standard Model gauge group $SU(3)_c \times SU(2)_L \times U(1)_Y$. Under
these assumptions, and ignoring any cubic and quartic self-interaction of $A$,
the part of the Lagrangian involving only the $\chi$ and $A$ fields can be
written down as
\begin{align}\label{eq:LDS}
{\cal{L}}_{\rm DS} = \frac{1}{2} (\partial^\mu A) (\partial_\mu A) - \frac{m_A^2}{2} A^2 + \frac{1}{2} \bar{\chi} \left(i\slashed{\partial}-m_\chi\right) \chi - i \frac{y_\chi}{2} A \bar{\chi} \gamma_5 \chi \, ,
\end{align}
where $y_\chi$ denotes the strength of the interaction of the mediator with
dark matter.
Being a singlet under the Standard Model gauge group, $A$ cannot couple to
quarks and leptons through renormalisable gauge-invariant interactions. Hence,
in order to parameterise the coupling of the dark sector with the visible one,
we introduce the effective Lagrangian
\begin{equation}\label{eq:LSM}
{\cal{L}}_{\rm f} = - i \sum_{f_u} c_u \frac{m_{f_u}}{v} A \bar{f}_u \gamma_5 f_u - i \sum_{f_d} c_d \frac{m_{f_d}}{v} A \bar{f}_d \gamma_5 f_d
\end{equation}
where the sums run over all up-type ($f_u$) and down-type ($f_d$) fermions respectively. The
$c_u$ and $c_d$ coefficients parametrise the strength of the interactions between $A$ and the SM fermions.
In the spirit of Minimal Flavour Violation, we moreover take the corresponding operators to be proportional to the ratio of the
SM fermion masses $m_{f_{u,d}}$ over the vacuum expectation value of the Higgs field
$v$. The Lagrangians of Eq.~\eqref{eq:LDS} and Eq.~\eqref{eq:LSM} will serve as
a basis for the analysis that follows.

In the above model description we have omitted, for the sake of simplicity, a
quartic term involving bilinears of the $A$ field and the SM Higgs doublet. Such
a term could have interesting phenomenological consequences for dark matter,
collider and Higgs phenomenology~\cite{Baek:2017vzd}, and also on first order electroweak baryogenesis~\cite{Ghorbani:2017jls}. However, this falls beyond the scope of the present work. 

It is also worth briefly commenting upon the potential UV origins of the Lagrangians \eqref{eq:LDS} and \eqref{eq:LSM} in order to motivate some of the parameter choices we will be adopting later on and to set the stage for the discussion that follows. The most straightforward UV completion of our setup would be in the framework of the two-Higgs doublet model (2HDM) or models involving even more extended scalar sectors. For example, in the context of type-II 2HDM, and denoting as usual by $\tan\beta = v_2/v_1$ the ratio of the two vacuum expectation values of the neutral CP-even components of the two Higgs doublets, we would obtain $c_u = \cot\beta$ and $c_d = \tan\beta$. However, in such a scenario it is the Lagrangian \eqref{eq:LDS} that would not be gauge-invariant. One solution could be to further introduce an additional scalar gauge singlet which mixes with the 2HDM Higgs doublets (an approach followed, \textit{e.g.}, in Ref.~\cite{Bauer:2017ota,Ipek:2014gua,No:2015xqa,Goncalves:2016iyg}) or, alternatively, to consider a pure 2HDM but extend the dark matter sector to a ``bino-higgsino'' or ``bino-wino''-like fermion system \cite{Banerjee:2016hsk,Bharucha:2017ltz}. 

In all these cases, additional interactions (\textit{e.g.} with extra scalars) would
arise at tree-level. Such interactions can introduce important phenomenological features which cannot be captured by the simple Lagrangians of Eq.~\eqref{eq:LDS} and Eq.~\eqref{eq:LSM}. However, depending on the specificities of each potential generalisation of our framework, these additional features can be extremely model-dependent, rendering generic statements extremely hard (if possible) to extract. In this spirit, we adopt the simpler description provided by the Lagrangians of Eq.~\eqref{eq:LDS} and Eq.~\eqref{eq:LSM}, cautioning the reader about subtleties that can appear in UV embeddings of our setup
(see section \ref{sec:discussion}).
Similarly, we ignore constraints that can potentially arise from precision
electroweak tests or flavour observables as any realistic assessment of their
impact would depend heavily upon the details of the UV embedding of the model.
This has been addressed, for instance, in the framework of the 2HDM, in the
work of Ref.~\cite{Haber:2010bw}. Corrections to the $S$, $T$ and $U$ oblique
parameters~\cite{Peskin:1990zt} or flavour constraints are nonetheless expected
to be subleading for the $A$ mass ranges under
consideration~\cite{Dolan:2014ska}. In this work, we thus focus solely on
{\it direct} probes of the dark matter particles $\chi$ and of the mediator $A$,
whilst keeping track of the limitations of our simplified framework.

From a low-energy standpoint, the couplings $y_\chi$, $c_u$ and $c_d$ in
Eq.~\eqref{eq:LDS} and Eq.~\eqref{eq:LSM} can take any numerical value (much
like the new physics masses $m_A$ and $m_\chi$) as long as perturbativity is respected. Throughout our study, results
are shown for different discrete combinations of three out of these five
parameters, the two others being varied freely. Our choices are mostly driven by
phenomenological considerations while keeping in mind some model-building
issues. In particular, we consider two distinct scenarios for the relative size
of the $c_u$ and $c_d$ coefficients, namely
\be
  c_u = c_d = 1 \quad\text{or}\quad
  c_u = c_d = 2 \ ,
\ee
which we refer to as ``top-dominated scenarios'' and 
\be
  c_u = 0.2, \qquad c_d = 20
\ee
that is referred to as a ``bottom-dominated scenario''. Interpreted, for example, in terms of a type-II 2HDM-like setup, the former case would correspond to $\tan\beta = 1$ with standard ($c_{u,d} = 1$) or somewhat enhanced ($c_{u,d} = 2$) Yukawa couplings, whereas the latter to $\tan\beta = 10$, again assuming slightly enhanced Yukawa couplings.
However, these values for $c_u$ and $c_d$ have mostly been chosen because, as
shown below, they allow for the exploration of different facets of the LHC
phenomenology associated with our model.

On the other hand, the dark matter relic abundance depends straightforwardly on the
mediator mass $m_A$. The latter is thus generically fixed when the
dark matter phenomenology of the model is addressed. Conversely, as the LHC
phenomenology of the model does not depend drastically on the dark matter mass
itself (up to phase space considerations), but rather on its relation to the
mediator mass, the discussion on the LHC constraints applicable to the model is
performed within two setups. Either all coupling values are fixed and we vary
the two new physics masses independently, or we
fix the dark matter mass $m_\chi$ and we vary its coupling to the mediator
$y_\chi$ along with $m_A$.

\section{LHC phenomenology}\label{sec:lhc}

In order to check the compatibility of our new physics
scenario with current LHC searches, we implement the model described in
section~\ref{sec:model} in the {\sc FeynRules} package~\cite{Alloul:2013bka} and
export it under the {\sc UFO} format~\cite{Degrande:2011ua} in order to make use of the {\sc
MadGraph5\_aMC@NLO} platform~\cite{Alwall:2014hca} for the simulation of hard-scattering
LHC collisions at centre-of-mass energies of 8, 13 and 14~TeV. When needed, these
fixed-order results are matched with parton showers within the {\sc Pythia 6}
environment~\cite{Sjostrand:2006za} that is also employed for describing the
hadronisation process, and we simulate the response of the LHC detectors with
the {\sc Delphes 3} software~\cite{deFavereau:2013fsa}.
When comparing our results with available experimental limits, we fold the cross
sections computed through {\sc MadGraph5\_aMC@NLO} with appropriate $K$-factors
to take into account non-simulated higher-order QCD corrections. For processes
in which the leading-order contribution to the scattering amplitude arises at
tree-level, as for the associated production of a top or bottom pair with dark
matter, the corresponding $K$-factors are computed directly using our model
implementation within the {\sc MadGraph5\_aMC@NLO} framework. In contrast, for
processes whose lowest order diagrams are already at the one-loop level, the
relevant $K$-factors are extracted from the literature, when available. These
issues are further elaborated upon in the following paragraphs.

In order to systematise the discussion, we divide the presentation of the various LHC constraints between those involving invisible decays of the pseudoscalar mediator and those where the mediator decays into visible Standard Model objects.

%%%%%%%%%%%%%%%%%%%%%%%%%%%%%%%%%%%%%%%%%%%%%%%%%%%%%%%%%%%%%%%%%%%%%%%%%%%%%%%%%%%%%%%%%%%%%%%%%%%%%%%%%%%%%%%%%%%%%%%%%%%%%%%%%%%%%%%%
\subsection{Invisible mediator decay}

%%%%%%%%%%%%%%%%%%%%%%%%%%%%%%%%%%%%%%%%%%%%%%%%%%%%%%%%%%%%%%%%%%%%%%%%%%%%%%%%%%%%%%%%%%%%%%%%%%%%%%%%%%%%%%%%%%%%%%%%%%%%%%%%%%%%%%%%
\subsubsection*{Monojet and multijet plus missing transverse energy searches}
\label{sec:monojet}

Monojet analyses are one of the primary probes for dark matter at the LHC, the
targeted signature being characterised by the presence of a hard QCD jet
recoiling almost back-to-back against a large amount of missing transverse
momentum. Although they were originally designed to veto events in which any
additional hadronic activity was present, it has been recently suggested that
allowing for the presence of extra jets could improve the sensitivity of these
searches, especially in the case where the partonic reaction is initiated by
gluon fusion~\cite{Buchmueller:2015eea}. For this reason, both the ATLAS and CMS
collaborations now include, in their monojet searches, signal regions populated
by events involving more than one hard jet~\cite{Aaboud:2016tnv,%
Sirunyan:2017hci}.

In this work we assess the compatibility of our scenario against the ATLAS
monojet search with 3.2~fb$^{-1}$ of integrated luminosity of proton-proton
collisions at a centre-of-mass energy of 13~TeV~\cite{Aaboud:2016tnv}. We use a
recasted version of this analysis~\cite{ma5-monojet} implemented in the {\sc
MadAnalysis 5} framework~\cite{Conte:2012fm,Conte:2014zja,Dumont:2014tja} and
available from the {\sc MadAnalysis} Public Analysis Database\footnote{
\url{https://madanalysis.irmp.ucl.ac.be/wiki/PublicAnalysisDatabase}.}, following the
procedure described in Appendix \ref{sec:mergingandmatching} for signal simulation.
This analysis targets monojet events and contains various signal regions
characterised by the considered amount of missing energy. Each region is
associated with a different $\slashed{E}_T$ selection threshold, the
hardest selection corresponding to $\slashed{E}_T> 700$~GeV. In order to
quantify the reach of such a monojet search for higher integrated luminosities,
which potentially opens the door to more aggressive $\slashed{E}_T$ thresholds,
we define three additional signal regions in which the missing energy
is required to be larger than 800, 1000 and 1200 GeV respectively. We extract
the corresponding Standard Model background expectation from the official ATLAS
$\slashed{E}_T$ distributions that cover a missing transverse energy range
extending up to 1200~GeV~\cite{Aaboud:2016tnv}, and define the related
uncertainty $\Delta N_{\rm bkg}$ on the number of expected background events
$N_{\rm bkg}$ as~\cite{Barducci:2016fue}
\be
  \Delta N_{\rm bkg}^2 = \Big(\kappa_1 \sqrt{N_{\rm bkg}}\Big)^2
    + \Big(\kappa_2 N_{\rm bkg}\Big)^2
  \quad\text{with}\quad
  \kappa_1=1.5 \ \text{and}\ \kappa_2=0.043 \ .
\ee
In this expression, the first term represents the statistical error and the
second the systematic one, the adopted values allowing us to adequately
parameterise the ATLAS results~\cite{Aaboud:2016tnv}.

We apply this analysis on
events for which the underlying matrix elements are allowed to contain up to
one extra jet and that are merged, after parton showering, following the `shower-$k_T$' merging scheme~\cite{Alwall:2008qv}. The first jet,
already present at the level of the lowest jet multiplicity matrix element, is
required to have a transverse momentum $p_T$ greater than 150~GeV to facilitate
the accumulation of a higher statistics in the analysed signal regions, given
the lowest experimental cut on the leading jet transverse momentum of 250 GeV.
Our study shows that limits obtained with 300 and 3000~fb$^{-1}$ of projected
integrated luminosity are actually comparable with those obtained through the
recasted version of the existing ATLAS analysis of 3.2~fb$^{-1}$ of data, this
lack of improvement being due to the $\sim4\%$ systematic uncertainty assumed in
the background determination, a number which is unlikely to improve in the
future. For this reason, we refrain ourselves from reinterpreting limits that
could originate from more recent LHC analyses, such as the CMS monojet search
with 12.9 fb$^{-1}$ of integrated luminosity~\cite{Sirunyan:2017hci} which
adopts a similar selection strategy and similar systematic uncertainty estimates
as in our projections.

Motivated by the fact that for a gluon fusion process a higher jet multiplicity
is expected in the final state, we have also examined the constraints that could
arise from a supersymmetry-inspired multijet ATLAS search~\cite{Aaboud:2016zdn},
basing our study on a recasted version of this analysis in the {\sc
MadAnalysis 5} framework~\cite{ma5-multijet}. The fundamental difference between
the monojet and multijet searches is that the latter involves a harder selection
on the additional jets. Eventually, it turns out that the reduction in signal
statistics outweighs the benefits of a more efficient background rejection, thus
leading to slightly weaker limits. Going a step further, we have checked whether
modifying a few selection cuts on the additional jets could improve upon the
sensitivity of the monojet analysis by means of the multivariate analysis
described in Appendix~\ref{sec:mva}. We have been unable to find any such
improved set of cuts, thus concluding that under their present form, monojet
searches appear to be optimal for targeting gluon-fusion-induced dark matter
production processes. As limits from multijet searches turn out to be subleading
with respect to the monojet-inspired ones, they are omitted in what follows.

%%%%%%%%%%%%%%%%%%%%%%%%%%%%%%%%%%%%%%%%%%%%%%%%%%%%%%%%%%%%%%%%%%%%%%%%%%%%%%%%%%%%%%%%%%%%%%%%%%%%%%%%%%%%%%%%%%%%%%%%%%%%%%%%%%%%%%%%
\subsubsection*{$t\bar{t}A$ and $b\bar{b}A$ searches, with $A \to \chi \chi$}\label{sec:ttabbaMET}

The associated production of the pseudoscalar $A$ with a pair of top or bottom
quarks, {\it i.e.} the same topology as for the production of the Standard Model
Higgs boson with a pair of third generation quarks, followed by the invisible
decay $A\to \chi\chi$, could be an efficient probe of our model. This is especially
true for top-dominated scenarios when the branching ratio
BR$(A\rightarrow \chi\chi)$ is substantial. Both the ATLAS and CMS
collaborations have performed searches for an invisibly decaying spin-0 mediator
particle produced in association with either a top or a bottom quark
pair~\cite{ATLAS-CONF-2016-050,CMS-PAS-EXO-16-005}\footnote{It has been shown in Ref.~\cite{Pinna:2017tay} that a pseudoscalar mediator produced in association with a single
top quark can yield stronger limits with respect to the conventional $t\bar{t}A,~A \to \chi \chi$ searches. However,
no current experimental data exists for such a search.}, so that these results can be
reinterpreted as constraints on the scenarios studied in this work.

We consider results from the ATLAS search for an invisibly decaying heavy
pseudoscalar mediator produced in association with a top quark pair in the
single-lepton plus jets plus $\slashed{E}_T$
channel~\cite{ATLAS-CONF-2016-050}. This search analyses 13.2~fb$^{-1}$ of
proton-proton collisions at a centre-of-mass energy of 13~TeV and exclusion
bounds at the $95\%$ confidence level (CL) are presented as contours in the
$(m_A, m_\chi)$ plane for a fixed and common choice of the mediator coupling to
the dark matter particle and to the top quark, $c_u = c_d = y_\chi = 3.5$. The
experimental publication moreover includes results for smaller values of this
common coupling setting in a restricted set of mediator and dark matter masses.
Although the number of points is too limited to draw an exclusion contour, it is
sufficient to check that this process does not constrain any further the 
parameter space of our model once other processes are accounted for, as
discussed in section~\ref{sec:results}.

To deduce limits from the ATLAS search for all relevant masses, we compute,
within the {\sc MadGraph5\_aMC@NLO} framework, the $t\bar{t}A$ associated
production cross section at the next-to-leading order (NLO) in QCD, and further
include leading-order (LO) branching ratios for the decays of all particles.
Furthermore, we consider the prospects of 300 fb$^{-1}$ of LHC collisions at a
centre-of-mass energy of 14 TeV. To this end, we import and reinterpret the
projected upper limits on the signal strength, computed with respect to
the case $m_{\chi}=1$~GeV and $c_u = c_d = y_\chi = 1$, from
Ref.~\cite{Haisch:2016gry}. Among the different investigations performed in this
last study, we choose the shape-based results that assume a 20\% uncertainty on
the background estimation.

Conversely, in bottom-dominated scenarios, it is instead the associated
production of the pseudoscalar mediator with a pair of bottom quarks that could
potentially provide the strongest constraints. The ATLAS and CMS collaborations
have performed analyses targeting this process for an integrated luminosity of
13.3~fb$^{-1}$~\cite{ATLAS-CONF-2016-086} and 2.17~fb$^{-1}$~\cite{CMS:2016uxr}
of LHC collisions at a centre-of-mass energy of 13~TeV, respectively. We make
use of their findings to compare, as for the $t\bar tA$ case, the predicted
signal cross sections with the excluded ones and derive exclusion
bounds on the parameter space of our model.

%%%%%%%%%%%%%%%%%%%%%%%%%%%%%%%%%%%%%%%%%%%%%%%%%%%%%%%%%%%%%%%%%%%%%%%%%%%%%%%%%%%%%%%%%%%%%%%%%%%%%%%%%%%%%%%%%%%%%%%%%%%%%%%%%%%%%%%%
\subsection{Visible mediator decay}
%%%%%%%%%%%%%%%%%%%%%%%%%%%%%%%%%%%%%%%%%%%%%%%%%%%%%%%%%%%%%%%%%%%%%%%%%%%%%%%%%%%%%%%%%%%%%%%%%%%%%%%%%%%%%%%%%%%%%%%%%%%%%%%%%%%%%%%%

In addition to  topologies where the mediator decays invisibly into a pair of dark matter particles, those involving decays into visible Standard Model states can be exploited to constrain our simplified model. 

\subsubsection*{$\tau^+\tau^-$ searches}\label{sec:tautaubar}

An important channel explored at the LHC is the production, 
either via gluon fusion or in association with a pair of bottom quarks, of a spin-0 mediator decaying 
into a pair of $\tau$ leptons. The CMS collaboration has performed an
analysis targeting these topologies with an integrated luminosity of
12.9~fb$^{-1}$ of LHC collisions at a centre-of-mass energy of 13~TeV. Final
states featuring either 0, 1 or 2 hadronically-decaying tau leptons have been
equally considered~\cite{CMS-PAS-HIG-16-037}, and the results have been
presented as 95\% CL upper limits on the production cross section of a heavy
Higgs boson in the context of the Minimal Supersymmetric Standard Model. Both
the gluon-fusion production channel and the associated $b\bar b A$ production
mode have been constrained, and we confront these results to the
predictions of our model. As no relevant information is provided explicitly, we assume
that interference effects with the Standard Model have not been considered, and
that the experimental results obtained for the case of a scalar mediator also
hold in the pseudoscalar case.

For the gluon fusion channel, we multiply the LO cross sections returned by {\sc
MadGraph5\_aMC@NLO} by N$^3$LO$_\textrm{A}$ + N$^3$LL$^{\prime}$
$K$-factors for pseudoscalar mediators~\cite{Ahmed:2016otz,Ahmed:2015qda}, so that the total rate includes
the matching of approximate next-to-next-to-next-to-leading order predictions
with the resummation of soft and collinear gluon radiation close to threshold
at the next-to-next-to-next-to-leading logarithmic accuracy. In the case of the
associated production of the mediator with a bottom quark pair, we consider
instead NLO production rates multiplied by LO branching ratios.

Although the production of a Higgs boson in association with zero, one or two
$b$-quarks and subsequently decaying into a pair of $b$-quarks or $\tau$-leptons 
has also been explored at the Tevatron, the related sensitivity does
not allow to probe pseudoscalar mediators lighter than
90~GeV~\cite{Aaltonen:2011nh,Abazov:2011up}. No additional constraints can hence
be deduced with respect to the ones extracted from the LHC results.

%%%%%%%%%%%%%%%%%%%%%%%%%%%%%%%%%%%%%%%%%%%%%%%%%%%%%%%%%%%%%%%%%%%%%%%%%%%%%%%%%%%%%%%%%%%%%%%%%%%%%%%%%%%%%%%%%%%%%%%%%%%%%%%%%%%%%%%%
\subsubsection*{$t\bar{t}$ searches}\label{sec:ttbar}
One of the strongest constraints that can be imposed on our simplified model,
especially in the case of top-dominated scenarios, comes from the study of new
physics effects that can potentially appear within the
production of a pair of top quarks. This corresponds to the $pp\to A^{(*)} \to
t\bar{t}$ process where the pseudoscalar can be either on-shell or off-shell.
While dedicated LHC searches for $t\bar t$ resonances deeply probe the on-shell
regime, the accurately measured $t\bar t$ production cross sections allow us to
constrain the new physics parameters in the case of off-shell production as well.

For the latter possibility we consider several studies in which the total
$t\bar{t}$ production cross section has been measured at the LHC, both for
centre-of-mass energies of 8 and 13~TeV. The most precise 13 TeV measurement
originates from the CMS analysis in the single-lepton plus jets channel, which
yields a $t\bar t$ production cross section of~\cite{Sirunyan:2017uhy}
\be 
  \sigma(t\bar{t})_{13} = 835 \pm 33~{\rm pb} \ ,
\ee
after summing in quadrature the various sources of uncertainties. The
corresponding Standard Model theoretical prediction, for a top mass
$m_t=172.5$~GeV, reads~\cite{twiki,Czakon:2011xx}
\be
  \sigma(t\bar{t})^{\textrm{theo.}}_{13} = 831.76^{+46.45}_{-50.85}~{\rm  pb} \ ,
\ee
with once again all sources of uncertainties added in quadrature. Assuming a
Gaussian spread for all uncertainties, we find a 95\% CL conservative upper
bound of
\be
   \sigma(t\bar{t})^{\textrm{NP}, 95\%{\rm CL}}_{13} = 120.43~{\rm pb}
\ee
on the size of the new physics contribution to the
$t\bar t$ total production cross section.

Similarly, the 8~TeV $t\bar{t}$ production cross section has been measured in
several channels by the ATLAS~\cite{Aad:2014kva,Aad:2015pga,Aaboud:2017rgh} and
CMS~\cite{Chatrchyan:2013faa,Khachatryan:2015fwh} collaborations. In order to
extract constraints, we have started from the ATLAS and CMS measurements in the
dileptonic mode~\cite{Aad:2014kva,Chatrchyan:2013faa},
\be
  \sigma(t\bar{t})_{8} = 242.9\pm 8.8~{\rm pb} \quad\text{and}\quad
  \sigma(t\bar{t})_{8} = 239\pm 13~{\rm pb} \ ,
\ee
that result in the same bound. We combine those numbers with the
Standard Model expectation for $m_t=172.5$~GeV~\cite{Czakon:2011xx,%
Czakon:2013goa},
\be
  \sigma(t\bar{t})^{\textrm{theo.}}_8 = 253^{+13}_{-15}~{\rm pb} \ ,
\ee
which finally enables us to extract an upper bound at the 95\% CL on the allowed
size for the new physics contributions,
\be
 \sigma(t\bar{t})^{\textrm{NP}, 95\%{\rm CL}}_8 \approx 25~{\rm pb} \ ,
\ee
all uncertainties being once again added in quadrature.

The 95\% CL upper bounds are then compared to predictions within our new physics
model. Interferences of the $pp\to A^{(*)} \to t\bar t$ contribution with the
Standard Model one should however be accounted for as they can be potentially
large. Technically, they cannot be computed directly by {\sc MadGraph5\_aMC@NLO} and a trick must be
employed. We first calculate the $ p p \to A^{(*)} \to t \bar{t}$ cross section
within our model and next match it to the one obtained in the context of a dummy
model where the heavy quark loop is approximated by a higher-dimensional
effective operator involving the mediator and gluons. Within this second model,
we then evaluate the interference with the SM $t\bar t$ background at LO, and we
multiply this result by a $K$-factor that we take as the geometrical mean
of the SM and new physics $K$-factors that are known separately. We find that
the interference effects are important. They rise with increasing $m_A$, are
maximal near the $t\bar{t}$ threshold, and decrease for larger mediator masses.
In particular, the interferences are often considerably larger than the new
physics contribution itself when the mediator mass lies right below the
threshold region. They must thus be included before evaluating the
constraints on the model.

We furthermore evaluate the potential constraints arising from resonance
$t\bar t$ searches. We use the Run~I ATLAS results obtained for an integrated
luminosity of 20.3~fb$^{-1}$ of 8~TeV collisions~\cite{Aad:2015fna}, assuming
that the quoted limits for a scalar mediator hold in our setup.

%%%%%%%%%%%%%%%%%%%%%%%%%%%%%%%%%%%%%%%%%%%%%%%%%%%%%%%%%%%%%%%%%%%%%%%%%%%%%%%%%%%%%%%%%%%%%%%%%%%%%%%%%%%%%%%%%%%%%%%%%%%%%%%%%%%%%%%%
\subsubsection*{Diphoton searches}

The pseudoscalar mediator $A$ couples to the ordinary fermions in a similar manner as the Standard Model Higgs boson, proportionally to their mass. This, in turn, implies that it can decay into a pair of photons through a loop involving top or bottom quarks, with the former contribution typically dominating over the latter unless $c_d \gg c_u$ as in our bottom-dominated scenario. The fact that the Standard Model Higgs boson was first observed in the $\gamma\gamma$ channel
motivates us to study constraints stemming from searches for diphoton resonances. The partial decay width of a pseudoscalar $A$ into a photon pair is given by \cite{Djouadi:2005gj}
\begin{equation}\label{eq:diphotonwidth}
\Gamma(A \rightarrow \gamma\gamma) = \frac{G_f \alpha^2 m_A^3}{128 \sqrt{2} \pi^3} \left| \sum_{f} N_c \ Q_f^2 \ c_f \ A^A_{1/2} \left(\tau_f\right) \right|^2 \ ,
\end{equation}
where $G_f$ and $\alpha$ are the Fermi and fine-structure constants
respectively, $N_c$ is the number of colours associated with each of the
fermions running in the loop, $Q_f$ their electric charge and the sum runs over
all Standard Model fermion species. Moreover, $\tau_f = m_A^2/4m_f^2$ and the
loop function $A^A_{1/2}$ is given by
\begin{equation}\label{eq:ggloopfunction}
A^A_{1/2} \left(\tau_f\right) = \frac{2}{\tau_f} f\left(\tau_f\right) \ ,
\end{equation}
with
\begin{equation}
f\left(\tau_f\right) = 
\begin{cases}
\arcsin^2\sqrt{\tau_f} & \tau_f \leq 1\ , \\
-\frac{1}{4} \left[ \log \left( \frac{1 + \sqrt{1-\tau_f^{-1}}}{1 - \sqrt{1-\tau_f^{-1}}} \right) -i\pi \right]^2 & \tau_f > 1 \, .
\end{cases}
\end{equation}

We employ the limits presented by the ATLAS collaboration in Ref.~\cite{ATLAS:2016eeo}
where 15.4~fb$^{-1}$ of 13~TeV collision data is used.
Based on the diphoton invariant mass distribution, this model-independent
analysis targets spin-0 resonances with masses as low as about 200~GeV, and the
results are presented as upper limits on the production cross section times
branching ratio for different choices of the resonance width. Similarly, we also
make use of the corresponding 8~TeV ATLAS results~\cite{Aad:2014ioa} that
extend the covered mass range down to about 65~GeV.

For the entire considered parameter range, we have verified that the total decay width of the mediator is always small enough so that the narrow width approximation holds.
This allows us to directly use the ATLAS limits and compare them
with our predictions at the N$^3$LO$_\textrm{A}$ + N$^3$LL$^{\prime}$ accuracy
for what concerns the mediator production rate, the corresponding branching
ratio into a photon pair being evaluated at LO. The interference effects with the continuum SM background have again been ignored here.

%%%%%%%%%%%%%%%%%%%%%%%%%%%%%%%%%%%%%%%%%%%%%%%%%%%%%%%%%%%%%%%%%%%%%%%%%%%%%%%%%%%%%%%%%%%%%%%%%%%%%%%%%%%%%%%%%%%%%%%%%%%%%%%%%%%%%%%%
\subsection{Other constraints}\label{sec:LEP_precision}

Several other collider searches could in principle also constrain the class of
dark matter scenarios under consideration. Amongst the visible mediator decay
processes, we have checked that mediator-induced dijet production does not
enforce any relevant constraint on the parameter
space~\cite{ATLAS-CONF-2016-070}, as do other mono-X searches such as the
monophoton ones~\cite{Aad:2014tda,Aaboud:2016uro}.
Another possibility are constraints arising from searches for the presence of
pseudoscalars in the decays of the SM Higgs boson, which could be relevant for
small enough mediator masses. Such searches have been performed, {\it e.g.}, in
the $4\tau$ channel~\cite{Khachatryan:2015nba,CMS:2015iga}. However, given the
restricted form of the scalar potential (see section~\ref{sec:model}), these limits do not apply in our case.

As regarding constraints arising from LEP data, searches were conducted for the
associated production of the mediator along with a pair of bottom quarks, with the subsequent $A\to b \bar b$ or $A\to \tau^+ \tau^-$ decay, and were used to derive constraints on pseudoscalars for masses lying in the [5, 50]~GeV window.
The obtained limits, at the 95\% CL, impose that
\be
  c_d \sqrt{{\rm BR}(A\rightarrow \tau\bar\tau)}< 12~~(80) \ ,
\ee
for $m_A=5$ (50) GeV in the $\tau^+ \tau^-$ final state and
\be
  c_d \sqrt{{\rm BR}(A\rightarrow b\bar{b})}< 20~~(100)\ ,
\ee
for $m_A=12$ (50) GeV in the $b\bar b$ final state~\cite{Abdallah:2004wy}.
Due to its larger branching ratio, the $b\bar{b}$ channel leads to the most
stringent upper limit on $c_d$, but the derived constraints are evaded as we
consider larger mediator masses not reachable at LEP.

Constraints from precision measurements on simplified models should be interpreted with care as additional contributions or cancellations can occur within a UV-complete theory. We therefore only  comment on  their potential impact.
Pseudoscalar contributions to the gauge boson self-energies appear at the
two-loop level and we therefore only expect weak constraints on the model from
precision electroweak measurements.
Constrains on light pseudoscalars ($m_A<10$ GeV) can be obtained from flavour physics, in  particular from measurements of the $B_s\rightarrow \mu^+\mu^-$ decay. Imposing the loose requirement  that the pseudoscalar contribution does not exceed the SM expectation leads to the bound~\cite{Dolan:2014ska} 
\be
  \sqrt{c_u c_d}<2 \frac{m_A}{10~{\rm GeV}} \ ,
\ee
which would thus only constrain scenarios involving a very light pseudoscalar.
Even a more aggressive treatment of this constraint, as could be obtained by
subtracting the SM contribution from the measured value~\cite{Aaij:2017vad}
while ignoring any interference effect, does not significantly reinforce it.

A light pseudoscalar could finally also contribute to the muon anomalous magnetic
moment. Within a pseudoscalar simplified model, explaining the observed $3\sigma$ deviation from the
SM expectation requires~\cite{Hektor:2014kga} that
\be
  c_d>50 \qquad\text{for}\qquad m_A<15~{\rm GeV}\ .
\ee
The full range of parameters considered in this work is thus not affected.

\section{Dark Matter phenomenology}\label{sec:darkmatter}

The observation of missing energy signals or the detection of a new resonance at
the LHC only provide little information concerning the cosmological relevance of
the underlying physics. In this spirit, we wish to confront the constraints
arising from LHC searches with those stemming from DM-related experiments,
enforcing in this way that the properties of the $\chi$ particle do not
challenge the cosmological and astrophysical observations. With large enough
couplings such as to yield observable rates at the LHC, our dark matter
candidate thermalises in the early universe and its present abundance can be
computed in the framework of a standard thermal freeze-out. As the LHC
phenomenology connected to $CP$-even and $CP$-odd scalar mediators is largely
identical, the dark matter observables are actually the ones that have dictated
our initial choice of considering pseudoscalar rather than scalar mediators.
$CP$-even mediators are indeed essentially excluded by the results of direct
detection experiments such as LUX~\cite{Akerib:2016vxi} once they are combined
with relic density requirements, under the condition that caveats like
the invocation of mechanisms such as entropy injection that dilutes the DM
abundance are ignored~\cite{Gelmini:2006pq}.

In the following subsections, we proceed to a brief description of the
experimental constraints used in this work. All dark matter observables
hereafter have been computed with the {\sc micrOMEGAs}
code~\cite{Belanger:2014vza,Belanger:2010gh,Belanger:2006is} that relies on
{\sc CalcHEP}~\cite{Belyaev:2012qa} model files obtained from the implementation
of the model in {\sc FeynRules}~\cite{Christensen:2009jx}.

%%%%%%%%%%%%%%%%%%%%%%%%%%%%%%%%%%%%%%%%%%%%%%%%%%%%%%%%%%%%%%%%%%%%%%%%%%%%%%%%
\subsection{Relic abundance}
The abundance of dark matter in the universe today has been precisely measured
by the Planck mission~\cite{Ade:2015xua}. At present, its central value reads
\be
  \Omega_{\rm DM}h^2=0.1187 \pm 0.0012\ ,
\ee
with the uncertainty including a $1\sigma$ variation. Given the smallness of
this uncertainty, we compute in what follows the abundance for every point of
the parameter space and, upon interpolation, show the contour corresponding to
the central value provided by Planck.

The predicted annihilation cross section $\sigma v$ where $v$ denotes the DM
velocity, and thus the relic abundance, scales straightforwardly with the
parameters of the model. For $2m_\chi\ll m_A$, the DM annihilates only to SM
fermions and
\be
  \left. \sigma v \right|_{2m_\chi\ll m_A} \propto y_\chi^2 c_f^2 \frac{m_\chi^2}{m_A^4} \ ,
\ee
neglecting all possible phase space suppressions and where $c_f^2$ consists of
a linear combination of $c_u^2$ and $c_d^2$. This region, as shown below, turns
out to be very constrained by LHC searches. On the other hand, in the $m_\chi>m_A$
region the $\chi\bar\chi\to AA$ annihilation channel opens up, 
with the corresponding cross section scaling instead as
\be
  \left. \sigma v \right|_{m_\chi>m_A} \propto y_\chi^4 v^2/m_\chi^2 \ .
\ee
As this channel exhibits a $p$-wave (velocity) suppression in the partial-wave
expansion of the cross section, its contribution to the total annihilation cross
section may be subdominant depending on the values of the other model
parameters, especially when the top-channel $\chi\bar\chi \to t\bar t$ is
kinematically open.

%%%%%%%%%%%%%%%%%%%%%%%%%%%%%%%%%%%%%%%%%%%%%%%%%%%%%%%%%%%%%%%%%%%%%%%%%%%%%%%%%%%%%%%%%%%%%%%%%%%%%%%%%%%%%%%%%%%%%%%%%%%%%%%%%%%%%%%%
\subsection{Indirect detection}

With direct detection constraints being largely inefficient to probe the
parameter space of our dark mater scenario, the main astrophysical constraints
originate from DM indirect detection and, in particular, from measurements of
the continuum gamma-ray spectrum from dwarf spheroidal galaxies, from searches
for spectral features at the Galactic Centre and from measurements of cosmic ray antiproton fluxes.

Dwarf spheroidal galaxies (dSphs) are dark matter-dominated objects, a property
which makes them very clean targets to test dark matter interactions with the
visible sector. The most recent search for gamma-rays in dSphs has been
performed by the Fermi-LAT collaboration~\cite{Ackermann:2015zua}, and
experimental bounds on the thermally averaged DM annihilation cross section
have been derived under the assumption of a 100\% annihilation into given
Standard Model channels. However, in our model, dark matter can annihilate into
any SM fermion pair, if kinematically allowed, with a specific weight $\omega$
that depends on the choice for the couplings $c_u$ and $c_d$ in
Eq.~\eqref{eq:LSM}. In order to account for the presence of several
annihilation modes contributing to the gamma-ray signal, we recast the
experimental bound on the thermally-averaged total annihilation cross section as
\be
  \langle\sigma v\rangle^{\rm max}(c_i,m_\chi) =
   \left(\sum_j\dfrac{\omega_j(c_i)}{\langle\sigma v\rangle^{\rm exp}_j(m_\chi)}
   \right)^{-1} \ ,
\label{svmax} \ee
where we sum upon all possible annihilation modes, where $c_i$ generically
stands for $c_u$ and $c_d$, and $\langle\sigma v\rangle^{\rm exp}_j$ are the
reported experimental 90\% CL limits on the annihilation cross section assuming
a 100\% annihilation into the channel $j$. The limit deduced from
Eq.~\eqref{svmax} is then compared with the $\langle\sigma v\rangle$ value predicted
by the model. As DM annihilation into a pair of
pseudoscalars is $p$-wave-suppressed, as mentioned above, this channel is always
subdominant and thus safely ignored from our analysis of the DM indirect
detection bounds.
In a similar fashion, we moreover report the expected 15-year exclusion bounds
that could be obtained assuming data issued from 60 dSphs, following the
projections of the Fermi-LAT collaboration~\cite{Charles:2016pgz}.
We, however, do not attempt to fit the Galactic Center excess observed by FermiLAT ~\cite{TheFermi-LAT:2015kwa,TheFermi-LAT:2017vmf}, which has been considered in the context of pseudoscalar mediated simplified models in Refs.~\cite{Boehm:2014hva,Kozaczuk:2015bea,Hektor:2017ftg} and EFT~\cite{Karwin:2016tsw}.

On different lines, cosmic ray antiprotons constitute at present one of the most
important channels for indirect dark matter searches. Recently, the AMS-02
experiment has released its results on the cosmic ray antiproton-to-proton ratio
with an improved statistical precision~\cite{Aguilar:2016kjl}, which has
recently allowed to derive an approximate $2\sigma$ limit on the
thermally-averaged DM annihilation cross section in the $b\bar b$
channel~\cite{Giesen:2015ufa}. We assume that this last limit
is representative for all quark flavours~\cite{Cirelli:2010xx} and then proceed to extract the maximum allowed
annihilation cross section according to the expression of Eq.~\eqref{svmax}.
The ensuing constraints however depend strongly on astrophysical assumptions such as the cosmic ray propagation model and the DM density profile. We use as a benchmark an Einasto DM halo profile \cite{Graham:2005xx}  and consider the so-called MED propagation model. As we will show in the following, for this benchmark the antiproton channel exclusion power is largely subdominant with respect to the sensitivity of the dSphs data. Given that dSphs constraints are generically considered to be more robust we will, thus, refrain from showing antiproton bounds adopting different astrophysical assumptions. We note, however, that as shown in \cite{Giesen:2015ufa}, opting for a MAX propagation model would amount to bounds on $\left\langle \sigma v \right\rangle$ which would be stronger by roughly one order of magnitude. A similar conclusion can be drawn from the recent analysis performed in \cite{Cuoco:2016eej}, with the exception of a dark matter mass range around 100 GeV where, depending on the whether or not the low-energy part of the AMS-02 antiproton data is taken into account, this group reports in general weaker limits. Besides, varying the assumed dark matter halo profile is known to only mildly affect the relevant constraints \cite{Giesen:2015ufa}. 

Finally, the coupling of the mediator $A$ to the Standard Model quarks can
induce, at the one-loop level, DM annihilation into a pair of photons that could
be observed as gamma-ray lines. The most stringent bounds on such a mechanism
are issued from the Fermi-LAT searches for spectral features at the Galactic
Centre~\cite{Ackermann:2015lka}. We estimate the related constraints on our
model by first matching the expression of Eq.~\eqref{eq:diphotonwidth} for the
mediator diphoton decay width to the one obtained when relying on an effective
interaction Lagrangian,
\be
  {\cal L}_{A\gamma\gamma} = \frac{\alpha}{4 \Lambda_\gamma} A
     \tilde{F}_{\mu\nu} F^{\mu\nu} \ ,
\ee
where $F$ and $\tilde F$ respectively stand for the photon field strength tensor
and its dual. The $m_A$ dependence of the form factor $A_{1/2}^A$ in
Eq.~\eqref{eq:diphotonwidth} is however
replaced by a $2m_{\chi}$ dependence, the latter being the relevant energy scale
of the process instead of the mediator mass. The DM annihilation cross section
into a photon pair is then computed in the resulting effective field theory,
with the scale $\Lambda_\gamma$ being appropriately fixed by the above
procedure.

%%%%%%%%%%%%%%%%%%%%%%%%%%%%%%%%%%%%%%%%%%%%%%%%%%%%%%%%%%%%%%%%%%%%%%%%%%%%%%%%
\subsection{Other constraints}

Similarly to indirect detection, Cosmic Microwave Background data can give rise
to {\it a priori} relevant constraints as well. Even if DM does not
significantly annihilate directly into electrons in our model, other highly
energetic annihilation products can heat and ionise the intergalactic medium,
although with less efficiency, and thus affect the last scattering surface which
is measured with high precision by the Planck collaboration. Taking the most
updated limits from Ref.~\cite{Kawasaki:2015peu}, we have verified that any
related constraint is roughly one order of magnitude weaker than those detailed
in the previous sections.

As already mentioned, the bounds on our model that can be obtained from DM
direct detection data are very weak as the effective operator that describes the
DM-nucleus interactions is momentum-suppressed. The current limits obtained by
the Xenon collaboration~\cite{Xenon_EFT} on several classes of operators (and in
particular the so-called operator $O_6$~\cite{Cheng:2012qr,Anand:2013yka} that
could be relevant in this work) do not further constrain the parameter space
under consideration.

%%%%%%%%%%%%%%%%%%%%%%%%%%%%%%%%%%%%%%%%%%%%%%%%%%%%%%%%%%%%%%%%%%%%%%%%%%%%%%%%
\subsection{Dark-matter-favoured regions of the parameter space}

Before closing this section and analysing the impact of LHC searches on our
model, we first determine which regions of the parameter space are favoured by
DM considerations. For given values of the mediator mass and of its couplings to
the Standard Model fermions, accommodating the correct relic density leads to an
$m_\chi$-dependent lower bound on the mediator coupling to dark matter $y_\chi$.
The strength of this coupling has to be large for light DM, is minimum for
$m_\chi\approx m_A/2$ and increases again for larger DM masses until
a new efficient annihilation channel opens up at the top-quark threshold
$m_\chi\sim m_t$. This in turn requires a small $y_\chi$ value to saturate the
Planck bound. Moreover, although we mainly restrict ourselves to the parameter
space region in which $m_\chi < m_A$, a new annihilation channel $\chi\chi\to AA$ opens once the
$m_\chi = m_A$ threshold is crossed, which may dominate the total DM annihilation depending on the parameters of the model. In this case, the relic density predictions become independent, if the narrow-width approximation holds, of the coupling
between the mediator $A$ and the Standard Model.

\begin{figure}
  \centering
  \includegraphics[scale=0.7]{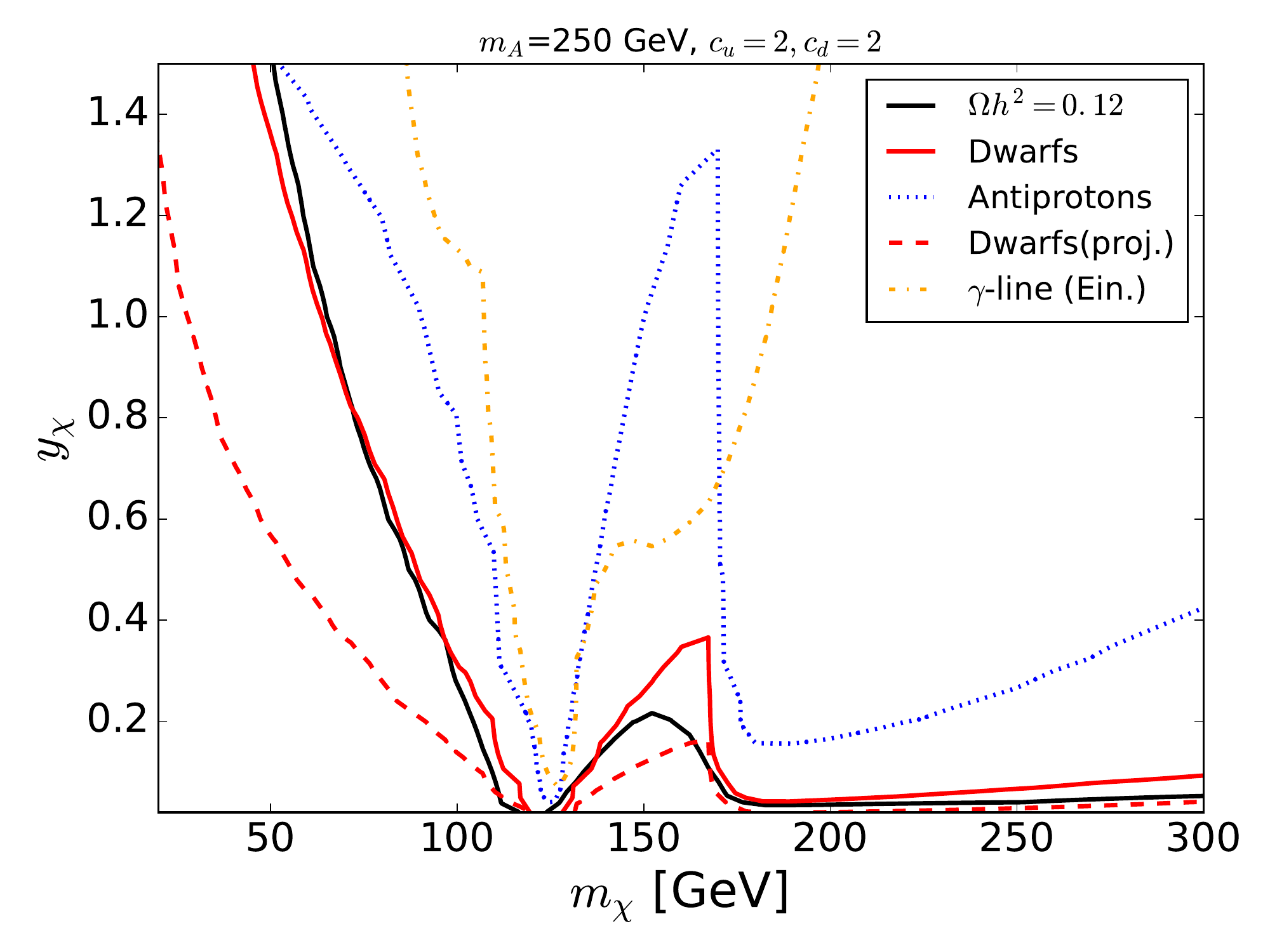} 
  \caption{\it Dark matter constraints on our model parameter space presented as
    contours in the $(m_\chi, y_\chi)$ plane when we assume a mediator mass
    $m_A=250$~GeV and $c_u=c_d=2$. In the region below the solid black line, the
    universe is overclosed whereas the regions above the solid red and dot-dashed yellow lines
    are respectively excluded by Fermi-LAT studies of the gamma-ray continuum in
    dSphs and of the gamma-ray lines in the Galactic Centre. The region above
    the dotted blue line is in addition disfavoured by AMS-02 antiproton measurements
    and the dashed red line represents the expected reach of Fermi-LAT after 15
    years of running.}
  \label{fig:MA_250}
\end{figure}

In Figure~\ref{fig:MA_250}, we present, as exclusion contours, all the 
constraints on our model that have been discussed in this section.
We fix the mediator mass to $m_A=250$~GeV and coupling strengths to the SM to
$c_u=c_d=2$, and show results in the $(m_\chi, y_\chi)$ plane. The various
combinations of parameters that can account for the entire dark matter abundance
in the universe according to standard thermal freeze-out are represented by a
black line. We can observe that the limits originating from the Fermi-LAT 
observations of dwarf spheroidal galaxies, shown as a solid red line,
are the most powerful ones for DM masses below the electroweak scale. These constraints
exclude the generic $s$-wave DM annihilation cross section favoured by Planck when DM
can only annihilate into light quarks or tau leptons. Consequently, the Fermi-LAT data
strongly disfavour the parameter space region compatible with the Planck lower bound for the relic density for
\mbox{$m_\chi<70$~GeV}. For
larger DM masses, the constraints become in contrast weaker and there
is always a set of couplings for which both Planck and dSphs constraints can be
satisfied simultaneously, although this region can be very narrow. Similar
results are found for other values of $c_u = c_d$ as long as $y_\chi \times c_f$
is kept constant. The exact location of the allowed region moreover depends on
the choice of the mediator mass, but a region allowed by all DM constraints can
always be found for pseudoscalar masses in the 10-1000 GeV range provided $c_u$
and $c_d$ are both not too suppressed. The projected 15-year Fermi-LAT limit,
depicted by a red dashed line, demonstrates that a conclusive statement about
the viability of the scenarios under consideration will be feasible. Almost the
entire parameter space compatible with Planck data is expected to be excluded,
except for a small resonant region around $m_\chi\approx m_A/2$, or if the DM particles are very heavy.

We also see from Figure~\ref{fig:MA_250} that constraints obtained from the
AMS-02 antiproton measurements (dotted blue line) are not as powerful as those
stemming from dSphs data for the entire considered mass range. They however
supersede the Fermi-LAT bounds for large couplings $y_\chi\approx 2$ and light
DM with $m_\chi<50$~GeV. These results are nonetheless not as robust
as they include a substantial amount of assumptions, in particular concerning
the cosmic-ray propagation model. Besides, constraints from
gamma-ray line searches are always subleading when compared to gamma-ray
continuum analyses, as the higher sensitivity does not compensate the
loop-suppression factor entering the DM annihilation cross section into photon pairs in our model. 
Given these findings, we consequently omit, in the following, all
constraints stemming from antiprotons and gamma-ray lines and focus instead
on the limits and projections originating from searches for the gamma-ray
continuum in the dSphs.

\section{Results and discussion}\label{sec:resultsanddiscussion}

Having presented the collider searches we consider in our analysis and sketched the parameter
space region that is favoured by dark matter observables, we now estimate the
impact of both the LHC and the astrophysical constraints on our model.

%%%%%%%%%%%%%%%%%%%%%%%%%%%%%%%%%%%%%%%%%%%%%%%%%%%%%%%%%%%%%%%%%%%%%%%%%%%%%%%%%%%%%%%%%%%%%%%%%%%%%%%%%%%%%%%%%%%%%%%%%%%%%%%%%%%%%%%%
\subsection{Results}\label{sec:results}

\subsubsection*{Fixed couplings}
\begin{figure}
  \centering
  \includegraphics[scale=0.7]{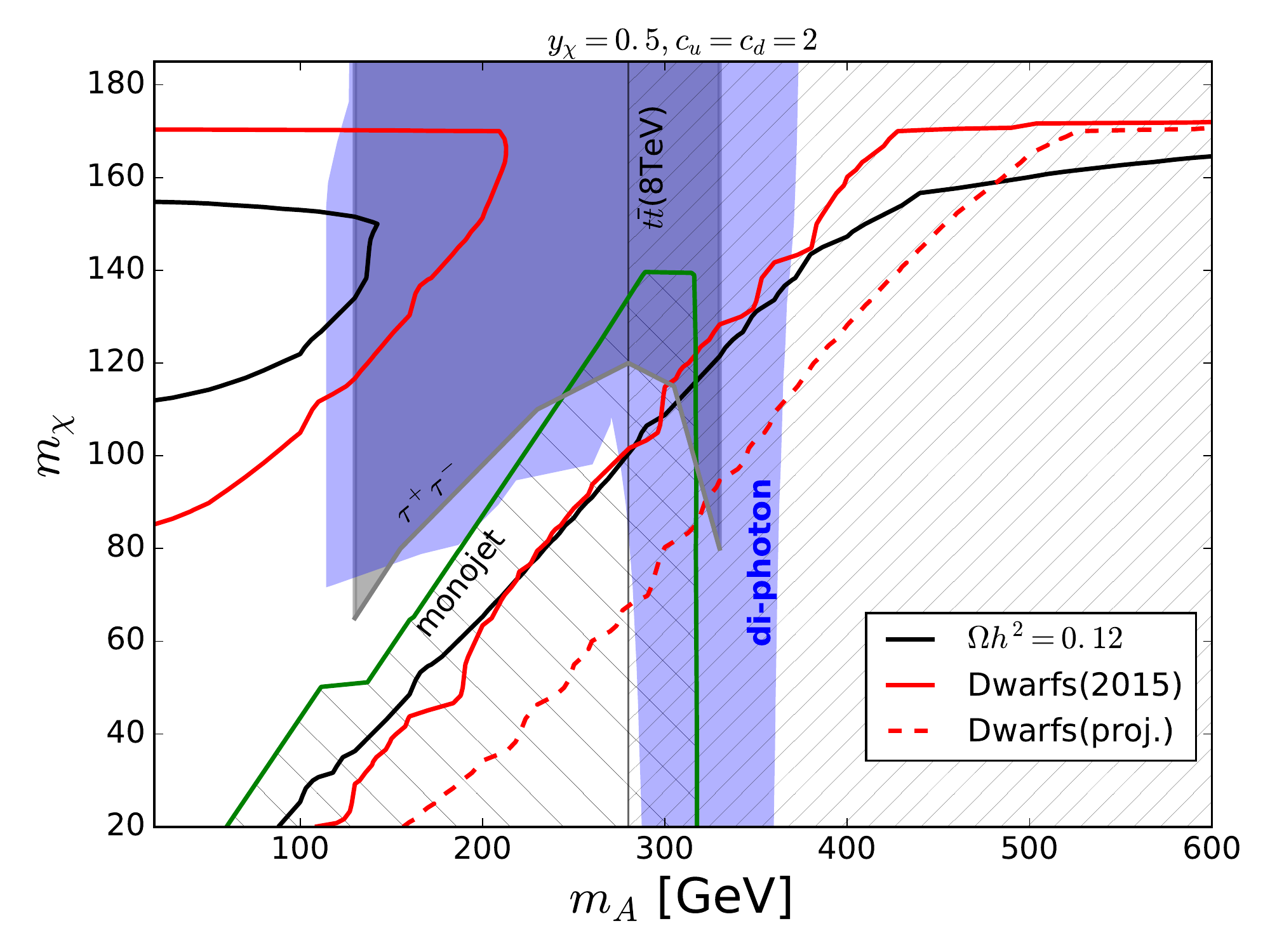} 
  \caption{\it Interplay of DM and collider constraints in the $(m_A, m_\chi)$
   plane for $y_\chi=0.5$ and $c_u=c_d=2$. The DM allowed region is located
   between the black and red line (see text for details). The shaded regions are
   excluded at the 95\% CL by LHC searches for monojets (hatched green),
   $A\rightarrow \tau^+\tau^-$ systems (grey), diphoton resonances
   (blue), and deviations in $t\bar{t}$ events at 8~TeV (hatched grey). }
\label{mass-mass}
\end{figure}

As a first step, following standard practice for the presentation of LHC dark matter searches \cite{Abdallah:2015ter,Abercrombie:2015wmb,Albert:2017onk}, we fix the mediator couplings to the visible and dark sectors and study the interplay of collider and DM observables in the
$(m_A, m_\chi)$ plane. LHC probes are more effective when the pseudoscalar
couples strongly to the Standard Model. In particular, we have found that the
$t\bar{t}$ searches almost single-handedly exclude the case  $c_u = c_d = 3$ for $m_A$ even below 100 GeV and up to more than 1~TeV. We therefore choose to show our results setting the couplings of the pseudoscalar to ordinary fermions to
lower values, picking as a benchmark scenario
\be
  c_u=c_d=2 \ ,
\ee
whereas we fix the coupling to dark matter particles to
\be
  y_\chi=0.5 \ .
\ee
Our results are shown in Figure~\ref{mass-mass}, where we highlight the $(m_\chi, m_A)$ combinations saturating the Planck bound (black solid lines), along with the parameter space regions excluded by the Fermi dSphs observations
(red solid lines) and 15-year projections (red dashed line), monojet searches
(hatched green region), $A\rightarrow \tau^+\tau^-$ searches (grey
region), diphoton resonance searches (blue region) and $t\bar{t}$ searches at
8~TeV (hatched grey region). 

There are two regions preferred from a dark matter standpoint. The first one corresponds to $m_A>300$ GeV (the area above the black and below the red lines towards the right-hand side in Figure~\ref{mass-mass}) and the other to lower values of  $m_A$ with $m_\chi\approx 150$ GeV (the area between the black and red lines towards the top-left of the same figure). For $m_A \lesssim 100$ GeV, dark matter is underabundant for small DM masses and the relic density increases with $m_\chi$ until it reaches the Planck limit around $m_\chi \sim 110$ GeV. Then, as soon as the $t\bar{t}$ annihilation channel opens up, the DM annihilation cross section increases significantly and $\chi$ becomes underabundant again. Besides, due to this sudden increase in $\langle \sigma v \rangle$, $m_\chi$ values above $m_t$ are excluded from Fermi-LAT. However, because of the finite velocity of dark matter particles in the early Universe, this increase occurs at a somewhat lower value of $m_\chi$ for the annihilation cross section during freeze-out (concretely, around $m_\chi \simeq 155$ GeV). This explains why dark matter masses around $160$ GeV are more difficult to probe with Fermi-LAT, as can be seen in Figure~\ref{mass-mass}: they correspond to situations where the observed dark matter abundance is obtained through annihilation into $t\bar{t}$ pairs in the early Universe, with this process being inefficient at present times. In any case, the 15-year projected limits of Fermi-LAT can fully exclude the parameter space regions of moderate pseudoscalar masses.

Because of the strong coupling to top quarks, the region with a heavy
pseudoscalar is severely constrained by the LHC. The 13 TeV $t\bar{t}$ total
cross section measurement excludes the considered model configuration for
pseudoscalar masses $m_A$ between 310~GeV and 410~GeV\footnote{We do
not show this constraint in the figures, first for the sake of clarity, and
secondly because it is always weaker than the constraints arising from a combination of the $t\bar t$ cross section measurements and $t\bar t$ resonance searches at 8 TeV.}, while the corresponding 8 TeV one further pushes the lower limit to $m_A=$ 280 GeV and the upper one to 430 GeV. The strongest constraint beyond $m_A = 400$ GeV comes from the 8 TeV $t\bar{t}$ resonance search which excludes the considered benchmarks for mediator masses ranging up to
$\sim800$ GeV. The process $At\bar{t}$ where $A$ decays invisibly does not provide additional constraints: we have found that only DM masses $m_{\chi} < m_A/2$ are excluded at the 95\% CL as long as $m_A<200$~GeV, while for $m_A> 300$ GeV all parameter space points are allowed\footnote{This constraint is not shown in
Figure~\ref{mass-mass} as only a few points are available from the experimental analysis, which are not sufficient to draw a meaningful contour.}.

The monojet and multijet constraints are only efficient for $m_A>2m_\chi$ and
their reach extends up to $m_A=320$ GeV, mostly probing Planck-preferred regions which are already challenged by Fermi-LAT. 
The searches for diphoton resonances also constrain the region  $100 \lesssim m_A \lesssim 350$~GeV. The lower value is due on one hand to the suppression of the diphoton branching ratio for small $m_A$ values and on the other hand to the reduced sensitivity of the LHC to light diphoton resonances. The upper value is a consequence of the fact that once the $t\bar{t}$ threshold is reached, the decay width of $A$ into top-quark pairs dominates and the diphoton channel constraints disappear. The non-trivial $m_A$-dependence of these limits in the intermediate region is due to a combination of two effects: generically, once the decay $A \rightarrow \chi\chi$ becomes kinematically allowed the branching ratio into photon pairs decreases substantially and the diphoton constraints vanish. However, as $m_A$ approaches $2 m_t$ the loop form-factor~\eqref{eq:ggloopfunction} entering the diphoton decay width~\eqref{eq:diphotonwidth} becomes maximal and the constraints become stronger, until the $A \rightarrow t\bar{t}$ channel opens up. Besides, direct searches for pseudoscalar decays in the $\tau^+\tau^-$ final state cover $m_A$ masses in the [120-320]~GeV range, provided the branching ratio into a ditau system is not too suppressed by the invisible partial width. In this channel, extending the search to lower masses will be required in order to probe the dark matter-allowed region for light pseudoscalars. We remind that all collider limits presented here correspond to a 95\% CL exclusion.

Reducing the coupling of the pseudoscalar to fermions (and in particular $c_u$) would considerably weaken the LHC constraints and especially the $t\bar{t}$ one as will be seen below. The limits from the diphoton channel also become weaker, since the loop-induced production cross section scales as $c_u^2$, with the bottom quark contribution to the process being subleading.

\subsubsection*{Fixed SM couplings and DM mass - benchmarks}
Next, we investigate the impact of the various constraints on the parameter
space for fixed values of the DM mass. We choose
\be
  m_\chi=100~{\rm GeV} \ ,
\ee
and, as discussed in section \ref{sec:model}, three sets of couplings as
\be
  c_u=c_d=2 \quad \text{(scenario {\bf S1})}
 \qquad\text{or}\qquad
  c_u=c_d=1 \quad \text{(scenario {\bf S2})}\ ,
\ee
which correspond to our top-dominated scenarios, as well as
\be
  c_u=0.2 \ \text{and}\  c_d=20\quad \text{(scenario {\bf S3})}\ ,
\ee
in the bottom-dominated scenario. The results are projected in the
$(m_A, y_\chi)$ plane.

\subsubsection*{Fixed SM couplings and DM mass - scenario S1}

\begin{figure}
  \centering
  \includegraphics[scale=0.7]{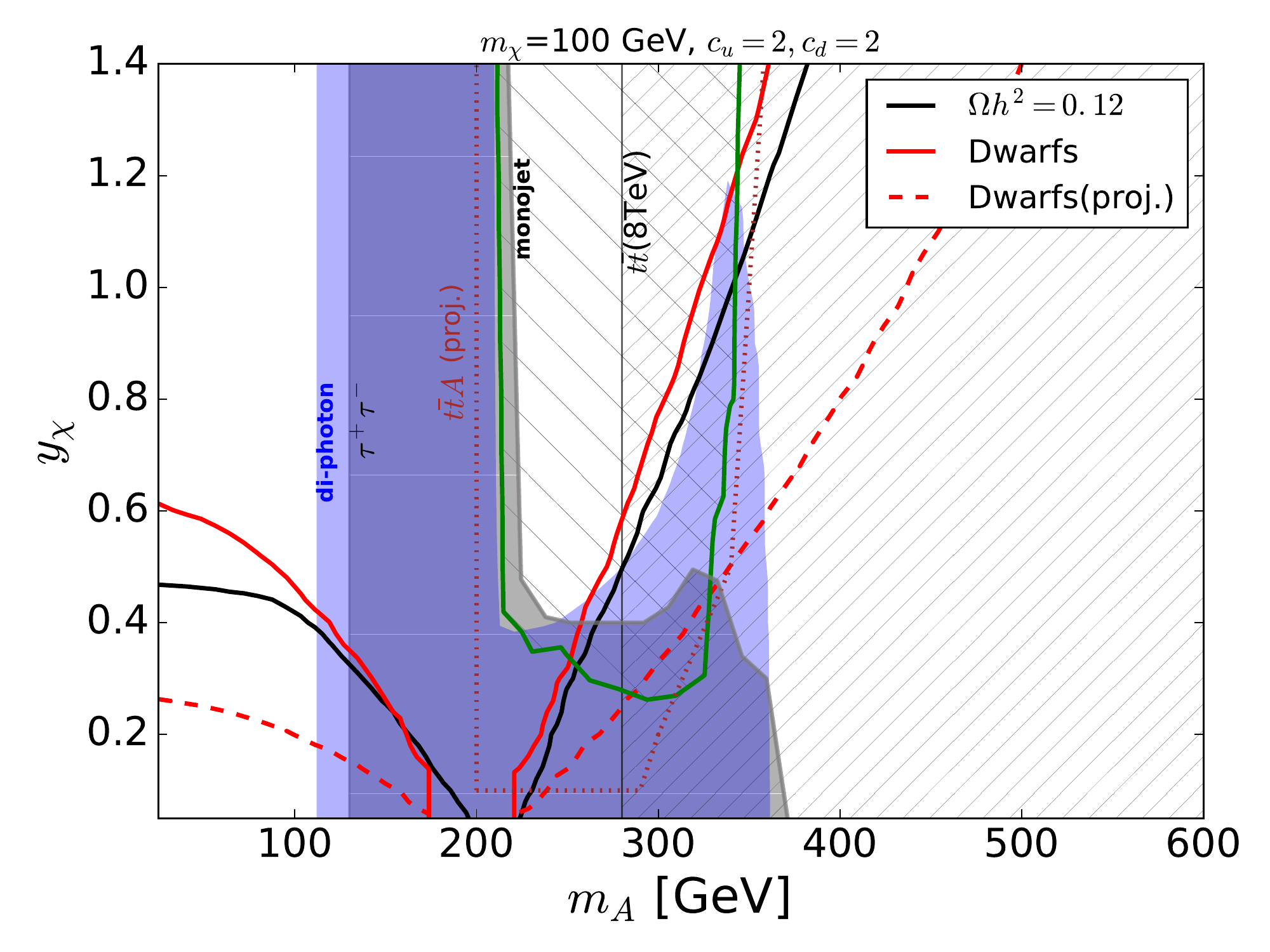} 
  \caption{\it Interplay of DM and collider constraints in the $(m_A, y_\chi)$
    plane for $m_\chi=100$~GeV and $c_u=c_d=2$ (scenario {\bf S1}). The
    region favoured by cosmology and astrophysics is located between the black
    and red line (see text for details). The shaded regions are excluded at the
    95\% CL by LHC searches for monojet (hatched green), $A\to\tau^+\tau^-$
    systems (grey), diphoton resonances (blue), and deviations in
    $t\bar{t}$ events at 8~TeV (hatched grey). Projections for $t\bar{t}A$
    probes for 300~fb$^{-1}$ of proton-proton collisions at 14~TeV (dotted red)
    and after 15 years of Fermi-LAT running (dashed red) are also displayed. }
\label{cu=cd=2}
\end{figure}

 We consider first the scenario {\bf S1} in which
$c_u=c_d=2$, for which our results are summarised in Figure~\ref{cu=cd=2}. As
mentioned before, $t\bar{t}$ searches place severe constraints on this scenario,
in particular due to the measurement of the $t\bar{t}$ cross section at 8~TeV
that entirely excludes the region $m_A>280$ GeV. We stress again that
interference effects are important here, and fairly independent of the value of
$y_{\chi}$.

The remainder of the DM-favoured region for $m_A> 2 m_\chi$ is also excluded by various searches. The $gg \rightarrow A\rightarrow \tau^+\tau^-$ search eliminates the region from $m_A=$ 130~GeV up to 220~GeV for all values of $y_{\chi}$. When the pseudoscalar can decay invisibly, its branching ratio into $\tau$ pairs is suppressed, inducing a  non-trivial dependence of the exclusion bounds on $y_{\chi}$. For  $y_{\chi}<0.5$, the limits extend to larger pseudoscalar masses reaching $m_A=$ 350 GeV for $y_\chi=0.3$. They sharply drop for  masses above  the $t\bar{t}$ threshold, where  no exclusion can be obtained from this channel.
These limits can potentially be extended to lower values of $m_A$, under the
condition the experimental results become publicly available.

Diphoton probes cover the region 120~GeV $ < m_A < $ 380~GeV, where the general behaviour of these constraints can be understood following a similar line of reasoning as before: the lower value is due to the suppression of the diphoton branching ratio for smaller $m_A$ and to the reduced LHC sensitivity. The upper value results from the same reduced production cross section combined with a reduced diphoton branching ratio given the large partial width into top pairs. Again, once $m_A$ becomes larger than $2 m_\chi$, a large value of $y_\chi$ implies an increased total width and a reduced branching ratio into photons, with the exclusion limits featuring a non-trivial dependence both on $y_{\chi}$ and on $m_A$. More precisely, as soon as the decay $A \rightarrow \chi\chi$
becomes kinematically allowed, we observe a sharp drop in the $y_\chi$ values
that are reachable due to the subsequent decrease in the diphoton branching
ratio. On the other hand, as $m_A$ increases further, the constraints become
stronger due to the maximisation of the loop form-factor of
Eq.~\eqref{eq:ggloopfunction} entering the diphoton decay width of
Eq.~\eqref{eq:diphotonwidth} for $m_A \sim 2 m_t$.

The monojet search possesses a similar sensitivity on the mediator mass, the
220~GeV $ < m_A <$ 340~GeV mass range being covered, but the results also
depend on $y_\chi$ as the search becomes ineffective when the pseudoscalar
coupling to DM is too small. In this sense, it is complementary to the diphoton
search.

The projection for 300~fb$^{-1}$ of future LHC collisions at a
centre-of-mass energy of 14 TeV shows that $t\bar{t}A$ probes (with
$A\to\chi\chi)$ will allow for the coverage of the parameter region currently
probed by monojet-like searches for mediator masses ranging up to
$m_A=350$~GeV and that the results depend on the value of $y_{\chi}$ below the
$t\bar{t}$ threshold. For smaller $y_{\chi}$ values of, for instance, 0.2, the
projected exclusion extends only to lower masses.

In summary, the only DM-favoured region that survives the various LHC bounds
lies below $m_A=130$~GeV and conclusive statements on the viability of this scenario will be achievable in the
future, thanks to the Fermi-LAT observations.

\subsubsection*{Fixed SM couplings and DM mass - scenario S2}
\begin{figure}
  \centering
  \includegraphics[scale=0.7]{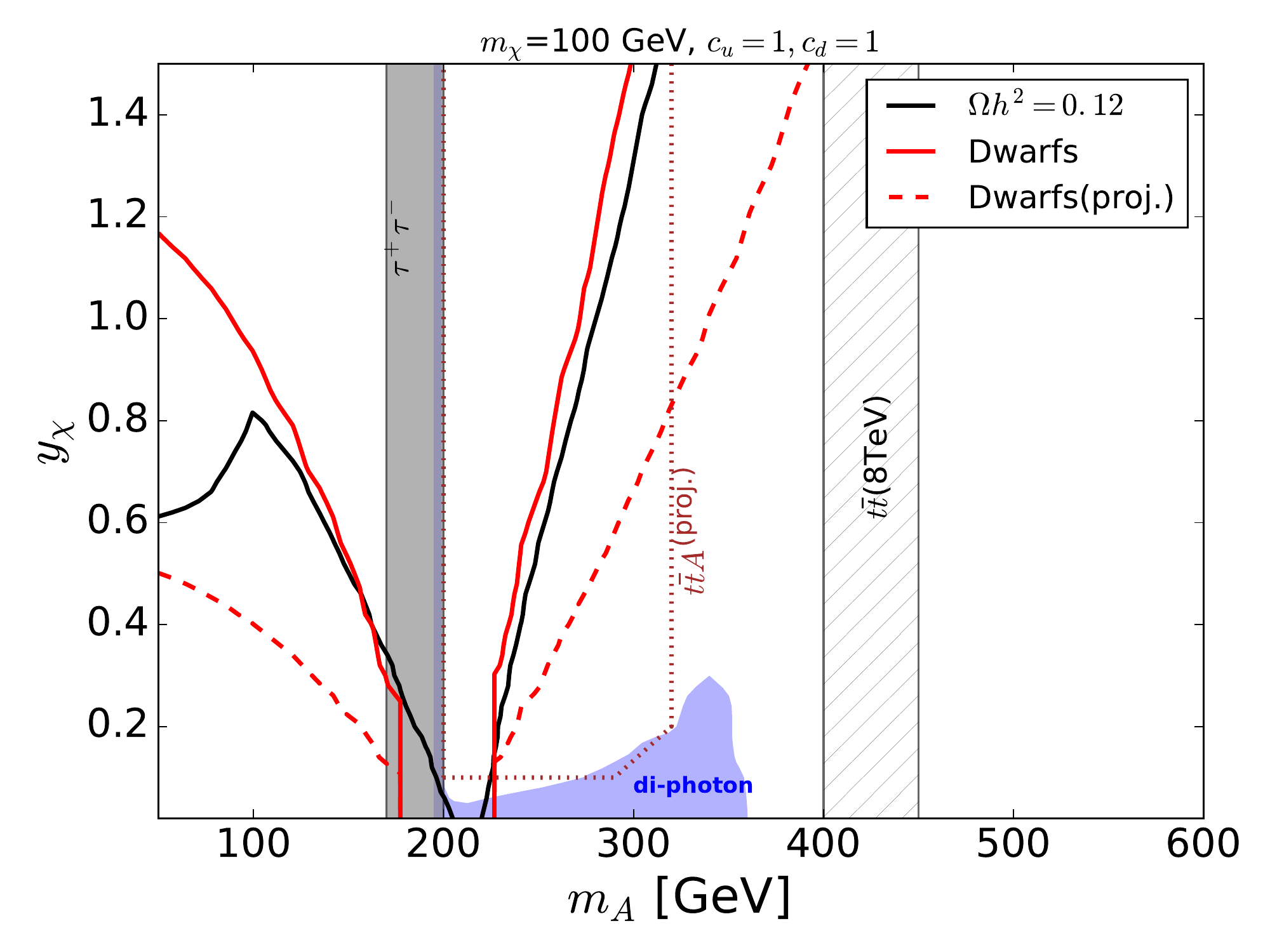}
  \caption{\it Same as Figure~\ref{cu=cd=2} but for scenario
    {\bf S2}.}
  \label{cu=cd=1}
\end{figure}
For weaker couplings to the SM particles, the LHC constraints get significantly
relaxed. The case $c_u=c_d=1$ (scenario {\bf S2}) is illustrated in
Figure~\ref{cu=cd=1}, where we for instance observe that searches with
$t\bar{t}$ probes become less efficient. There is only a small corner
of the {\bf S2} parameter space, in which 400~GeV $ < m_A <$ 450~GeV, that is
excluded by 8 TeV $t\bar t$ resonance searches. In addition, configurations for
which $m_A\in [170, 200]$~GeV are excluded by $p p \to A \to \tau^+ \tau^-$
searches independently of $y_{\chi}$ whereas the diphoton searches cover scenarios
in which $m_A\in [200, 380]$~GeV provided that $y_\chi< 0.2-0.3$, with the exact value 
depending on the mediator mass. Finally, the monojet
search in contrast barely excludes a very narrow mass range around $m_A=200$~GeV
for coupling $y_\chi\gtrsim 1.3$, which is not shown in the Figure. In summary, most of the {\bf S2} parameter
space region favoured by DM (between the black and red lines) is consistent with
all current LHC constraints. The mass region 200~GeV $< m_A <$ 320~GeV is
expected to be covered by the future results of the $t\bar{t}A$ LHC analyses of
14~TeV collisions, as long as $y_\chi>0.1$. 
Fermi-LAT will, besides, be able to exclude essentially the entire parameter space with 15 years of data acquisition.

\subsubsection*{Fixed SM couplings and DM mass - scenario S3}

\begin{figure}
  \centering
  \includegraphics[scale=0.6]{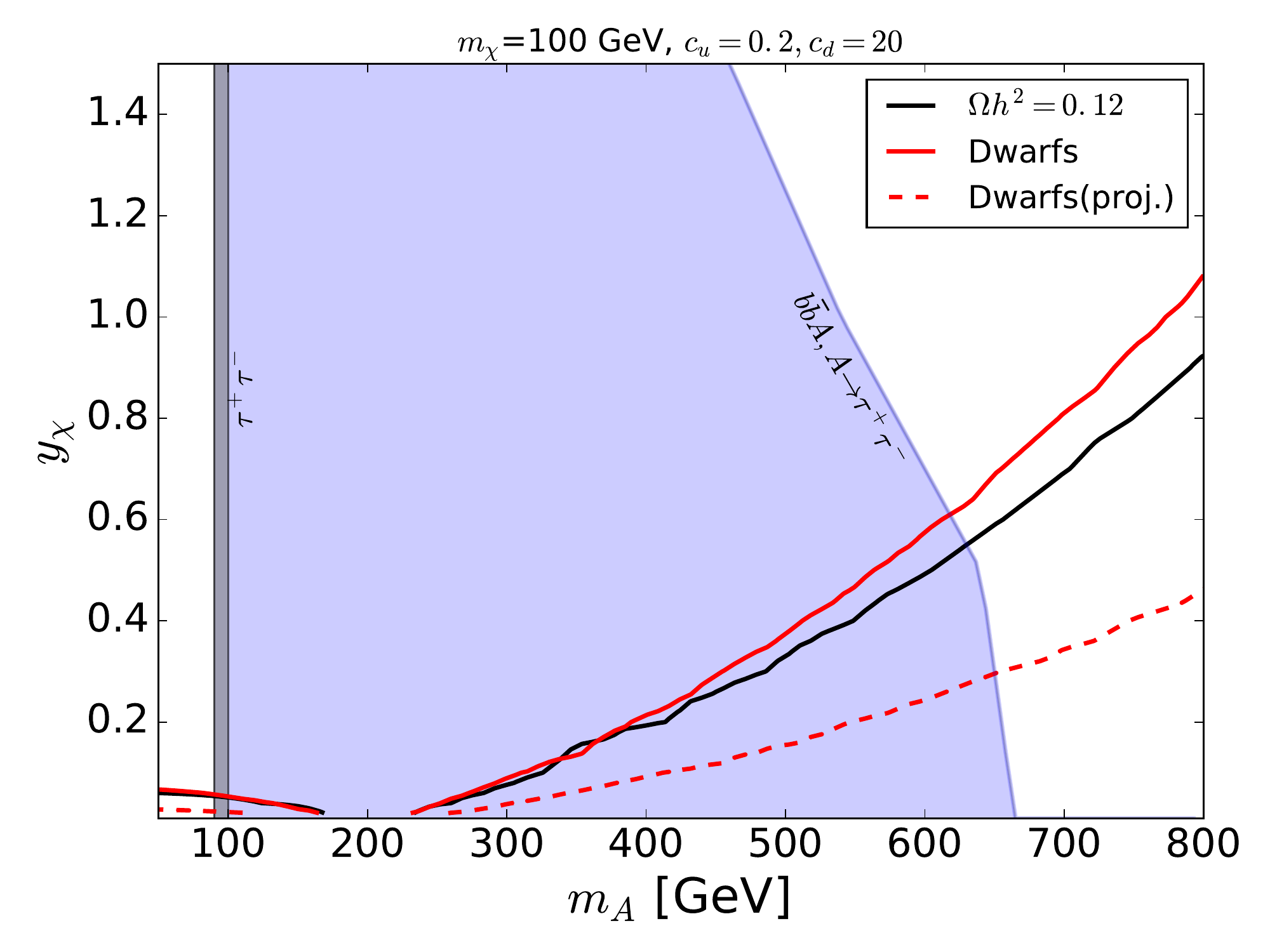} 
  \caption{\it Interplay of DM and collider constraints in the $(m_A, y_\chi)$
    plane for $m_\chi=100$~GeV and $c_u=0.2$ and $c_d=20$ (scenario 
    {\bf S3}). The DM allowed region is located above the black (Planck) and
    below the red (Fermi-LAT) lines. Shaded regions are excluded at the 95\% CL
    by LHC searches in the $b\bar{b}A,~A\rightarrow \tau^+\tau^-$ (blue) and
    $A\rightarrow \tau^+\tau^-$ (grey) modes. Projections for Fermi-LAT after
    15 years of running (dashed red) are also displayed.}
\label{cu=02cd=20}
\end{figure}
We finally consider a case in which the coupling to $b$-quarks is enhanced and
the one to $t$-quarks is suppressed (scenario {\bf S3}). As for the other
considered cases, we have found two viable DM regions, one for $m_A>350$~GeV and
another (very narrow) one for $m_A<90$ GeV. Clearly, no limits can be derived from
the $t\bar{t}$ or $t\bar{t}A$ processes. The main LHC constraints instead come
from searches in the $b\bar{b}A$ with $A \to \tau^+ \tau^-$ channel. Pseudoscalar masses between $m_A=$ 90 GeV and 450 GeV are fully excluded for $y_{\chi}<1.5$. The exclusion bounds extend to larger pseudoscalar masses but for smaller dark matter-mediator couplings, reaching up to $m_A=660$~GeV for $y_{\chi} \lesssim 0.1$. There is also a narrow excluded region between $m_A=$ 90 GeV and 100 GeV from the pseudoscalar (produced in the gluon fusion mode) search in $\tau^+\tau^-$. Again, the absence of data below $m_A=$ 90 GeV makes it difficult to predict the viability of the considered
scenarios in this region. Additional constraints could  be obtained from the search for $Ab\bar{b}$ with $A$ decaying to a pair of dark matter particles.
The limit extracted from Ref.~\cite{CMS:2016uxr} for $m_\chi=100$~GeV is not available, however it should be weaker than the quoted limit for $m_\chi=1$ GeV, for a given $m_A>2m_\chi$. We found that the regions excluded by this constraint would lie in the range 200 GeV$<m_A<$ 300~GeV for $y_\chi\gtrsim 1.5$ and are, thus, anyway covered by the  $b\bar{b}A,~A \to \tau^+ \tau^-$ searches. In summary, the DM-favoured regions in the bottom-dominated scenario survive the LHC constraints for $m_A>620$~GeV and $m_A<90$~GeV. As for our other scenarios, we expect both these regions to be probed by Fermi-LAT in the future.

%%%%%%%%%%%%%%%%%%%%%%%%%%%%%%%%%%%%%%%%%%%%%%%%%%%%%%%%%%%%%%%%%%%%%%%%%%%%%%%%%%%%%%%%%%%%%%%%%%%%%%%%%%%%%%%%%%%%%%%%%%%%%%%%%%%%%%%%
\subsection{Discussion}\label{sec:discussion}

In the previous subsection we presented results fixing three out of the five free parameters of our model at a time. In order to ascertain the generality of these results, we comment below on the behaviour of the limits for different choices of the dark matter mass and mediator couplings. On general grounds, LHC
constraints are mostly important for large couplings of the pseudoscalar to top quarks, at least for the top-dominated scenarios \textbf{S1} and \textbf{S2}.

As illustrated in Figure~\ref{fig:MA_250}, DM masses below $\sim 70$ GeV are
strongly constrained from a combination of Fermi-LAT and Planck results for
fixed (and actually rather large) values of the couplings $c_u$ and $c_d$. Moreover, decreasing the DM mass generically shifts the Planck-compatible region towards higher values of $y_\chi$ pushing it quickly into the large coupling regime. As for LHC constraints, both the monojet and $t\bar{t}A$ exclusion bounds are expected to be extended to lower values of $m_A$ (down to $m_A \sim 2 m_\chi$).
Alternatively, increasing the DM mass shifts the Planck-favoured region associated with $m_A>2m_\chi$ towards smaller couplings $y_\chi$ for a fixed pseudoscalar mass, and therefore this region  is constrained mainly from the $t\bar{t}$ channel (see {\it e.g.} Figure~\ref{cu=cd=2}). Other constraints are mostly limited to the region with $m_A< 2 m_t$ and are largely independent of the other parameters of the model -- for example, the precise value of the DM mass will only shift the lower limit of the exclusion bounds from invisible channels, which is anyway confined to $m_A>2m_\chi$. Finally, the limits coming from the Fermi-LAT dSphs searches become weaker for heavier dark matter, hence, the DM-favoured region will become more extended as $m_\chi$ increases. 

The results for the bottom-dominated case show that the main LHC constraints do not involve invisible final states, and are therefore mostly insensitive to the DM mass. However, the limits originating from $A\rightarrow \tau^+\tau^-$
searches will become stronger when the $A\to\bar\chi\chi$ decay is not kinematically open, thus covering the low $m_A$ values which are missed by the $bbA,~A\rightarrow \tau^+\tau^-$ searches. For example, we have checked that increasing the value of $c_d$ to 30 would amount to an exclusion of pseudoscalar masses up to $m_A \sim 860$ GeV. 

As a final remark, we should stress again the limitations of simplified models. In particular, the relic density predictions can vary substantially in concrete UV completions of our model due to the presence of additional annihilation channels involving, for example, extra scalars\footnote{For simplicity here we do not consider $CP$ violation effects, which could give rise to annihilation channels involving massive vector bosons.}. Such additional contributions to DM annihilation in general affect the relic density and the gamma-ray flux in the same direction and, hence, the parameter space regions allowed by both Planck and Fermi-LAT are expected to remain narrow. The main exception to this rule is connected
to co-annihilations, which only affect the relic density and therefore reduce the region where DM is overabundant. As for LHC limits, for $m_A < 2 m_t$, the presence of additional decay channels can indeed substantially modify our results. As long as $m_A > 2 m_t$, we however expect the limits obtained in this work to remain qualitatively valid (provided that the narrow-width approximation holds). Exceptions do exist, for instance in scenarios where additional contributions to the $t\bar{t}$ total cross section interfere with the SM and/or pseudoscalar ones. The situation becomes, of course, more involved once we consider the evolution of the interplay between cosmological/astrophysical constraints and collider ones, a case in which the hierarchy between all masses involved in potential UV completions of our simplified model should be taken into account.

\section{Summary and outlook}\label{sec:outlook}

From a combination of searches for particles decaying into Standard Model and/or invisible final states, we have shown that the LHC provides strong constraints on models of fermionic dark matter coupled to the visible sector through a pseudoscalar mediator $A$. The constraints apply mainly when at least one of the mediator couplings to quarks is of ${\cal O}(1)$ and cover a substantial fraction of the DM-favoured region of the parameter space. 

When the mediator possesses a sizeable coupling to top quarks, LHC dark matter searches based on invisible channels are only relevant for $m_A \lesssim 350$~GeV as for higher masses its total width is dominated by the decay into $t\bar{t}$. Direct searches for the mediator in channels involving top quark pairs overcome this restriction and, as we have shown, allow the LHC to probe pseudoscalar masses up to $800$ GeV, depending on the exact value of the relevant couplings. A dedicated search for resonances in the $t\bar{t}$ channel with increased luminosity is, thus, expected to significantly extend the LHC reach to larger mediator masses and to smaller values of its couplings to quarks. Moreover, searches for an invisibly decaying pseudoscalar produced in association with top quarks will be able to probe a significant fraction of the DM-favoured region when the mediator mass is in the range $2m_\chi<m_A<2m_t$. Our findings further show that the projected limits obtained from the $t\bar{t}A$ channel are complementary to the bounds stemming from searches for diphoton resonances as well as to the corresponding bounds from traditional monojet searches. Besides, pseudoscalar masses below 100-200 GeV survive all constraints, where the exact value depends on the choice of the DM mass and of the mediator couplings. The reason is either a lack of data in the region of interest, {\it e.g.}~in the ditau channel, or a suppressed branching ratio of the mediator, {\it e.g.}~in the case of diphotons. Thus, most of the DM-favoured regions at mediator masses of ${\cal O}(100)$~GeV or lower remain allowed once LHC constraints are imposed. However, if the mediator couples strongly to down-type fermions (in the so-called ‘bottom-dominated scenario’, {\bf S3}), this mass
range could be probed at the future LHC runs through searches for light pseudoscalars in the $b\bar{b}A$ channel, with the mediator decaying into a pair of taus. The light pseudoscalar scenario can also easily be probed at a future electron-positron collider in either the $\tau^+ \tau^-$ or $b\bar{b}$ mode,
as for such masses the associated cross section exceeds 20~fb~\cite{Kanemura:2014dea}. 
\\
\\
On the side of indirect detection, searches for gamma-rays from Dwarf Spheroidal galaxies will allow Fermi-LAT to cover most of the DM-favoured region after 15 years of data acquisition. One known exception is the DM resonance region where $m_\chi\approx m_A/2$.
In this case, the mediator couplings either to dark matter or to the Standard Model particles (or, eventually, to both) have to be very suppressed in order to saturate the relic density bound. In a narrow mass range, the total thermally averaged annihilation cross section $\left\langle\sigma v\right\rangle$ at galactic velocities indeed turns out to be much suppressed compared to $\left\langle\sigma v\right\rangle$ in the early universe, rendering even future indirect detection observations irrelevant. Although it could be argued that this parameter space region is fine-tuned, it generically occurs in all models where DM annihilation proceeds through an $s$-channel resonance. The LHC, however, can (at least partly) probe the DM resonance region through searches for the mediator decaying into visible final states such as top or tau pairs, provided its couplings with the Standard Model particles are not suppressed. This feature is yet another illustration of the complementarity between astroparticle and collider searches for dark matter.

%=========================================================================================
\section*{Acknowledgements}
%=========================================================================================
We thank Fawzi Boudjema, C\'edric Delaunay and Alberto Zucchetta for useful
discussions, and are also grateful to Laurent Duflot and Takashi Yamanaka for
discussions on the ATLAS search in the multijet plus missing energy channel, to
Priscilla Pani for providing us with digitised information for the ATLAS $t\bar t A$
search, to Valentin Hirschi for his valuable help with {\sc MadGraph5\_aMC@NLO},
to Shilpi Jain and Satyaki Bhattacharya for help with multivariate analyses,
to Narayan Rana for clarifications
about $K$-factors adopted for gluon-fusion processes and to Nicolas Berger for
ascertaining the fact that a combined Gaussian uncertainty is conservative even
if the actual distributions deviate from exact Gaussian shapes. This work was
supported in part by the French ANR project DMAstro-LHC (ANR-12-BS05-0006), by
the {\it Investissements d'avenir} Labex ENIGMASS, by French state funds managed
by the Agence Nationale de la Recherche (ANR) in the context of the LABEX ILP
(ANR-11-IDEX-0004-02, ANR-10-LABX-63), and by the Research Executive Agency
(REA) of the European Union under the Grant Agreement PITN-GA2012-316704
(HiggsTools). SB acknowledges support from the Indo-French LIA THEP (Theoretical
High Energy Physics) of the CNRS.

%=========================================================================================
\appendix
%=========================================================================================
\section{Merging and matching}\label{sec:mergingandmatching}

In order to accurately simulate the monojet and multijet processes described in
section~\ref{sec:monojet}, we have generated hard scattering events for the
$pp\to A$ process with matrix elements containing up to one extra jet. These
fixed-order results have been matched with the parton shower infrastructure of
{\sc Pythia 6} and then combined following the `shower-$k_T$' merging
scheme~\cite{Alwall:2008qv}. The merging parameters $Q^{\rm cut}$
(at the matrix-element level) and the merging scale $Q$ have been fixed to the
common $m_A$-dependent value reported in Table~\ref{tab:matching}.

We have moreover verified the stability of our results with respect to the case
in which up to two extra jets are allowed at the matrix element level. Due do
the prohibitive computational cost of performing this task with the full model
described in section~\ref{sec:model}, we have opted for a simpler effective field theory model described by the Lagrangian
\begin{align}\label{eq:fscpo}
  {\cal{L}}_{\rm EFT} & = \frac{1}{2} (\partial_\mu A)(\partial^\mu A) -
     \frac{m_A^2}{2} A^2 
  - \frac{g_1^2}{4 \pi} \frac{1}{4 \Lambda_1} A ~ B_{\mu\nu} \tilde{B}^{\mu\nu}
  - \frac{g_2^2}{4 \pi} \frac{1}{4 \Lambda_2} A ~ W_{\mu\nu} \tilde{W}^{\mu\nu} \\ \nonumber
 &- \frac{g_3^2}{16 \pi \Lambda_3} A ~ G_{\mu\nu} \tilde{G}^{\mu\nu}
  + \frac{1}{2} \bar{\chi} \left(i\slashed{\partial}-m_\chi\right) \chi - i \frac{y_\chi}{2} A \bar{\chi} \gamma_5 \chi \, ,
\end{align}
where $B_{\mu\nu}$, $W_{\mu\nu}$ and $G_{\mu\nu}$ are the $U(1)_Y$, $SU(2)_L$
and $SU(3)_c$ field strength tensors respectively, and $\tilde{B}_{\mu\nu}$,
$\tilde{W}_{\mu\nu}$ and $\tilde{G}_{\mu\nu}$ their duals. Moreover, $g_1$,
$g_2$ and $g_3$ are the hypercharge, weak and strong coupling constants. We have
verified that the two procedures yield comparable results once the experimental
selections for the monojet and multijet analyses are applied for a selected set
of scenarios, justifying our choice of simulating hard scattering events with up
to only one jet in the matrix element.

\begin{table}
\begin{center}
\begin{tabular}{ c | c }
  $m_A$ [GeV]        &  $Q^{\rm cut} = Q$ [GeV] \\
  \hline
  $50-100$           & 15  \\  
  $101-125$           & 25  \\
  $126-250$           & 35  \\
  $251-275$           & 40  \\
  $276-350$           & 50  \\
  $351-450$           & 60  \\
   $451-500$          & 70  \\
\end{tabular}
\caption{\it Parameters used on the multiparton matrix element merging procedure in
  the case of the $pp\to A$ process with matrix elements containing up to one
  extra jet.}
\label{tab:matching}
\end{center}
\end{table}

%%%%%%%%%%%%%%%%%%%%%%%%%%%%%%%%%%%%%%%%%%%%%%%%%%%%%%%%%%%%%%%%%%%%%%%%%%%%%%%%%%%%%%%%%%%%%%%%%%%%%%%%%%%%%%%%%%%%%%%%%%%%%%%%%%%%%%%%
\section{MVA analysis}\label{sec:mva}

Having employed the ATLAS monojet-like analysis in order to constrain our model,
we further attempted to check if any kind of improvement is possible in
this regard. To this end, we again relied on the EFT scenario described by
the Lagrangian of Eq.~\eqref{eq:fscpo}. 
As discussed in section~\ref{sec:monojet}, the ATLAS monojet-like analysis at
13~TeV relies on a strategy that requires the presence of a hard jet with a
transverse momentum $p_T > 250$~GeV and of at most
four additional jets with $p_T > 30$~GeV. Seven inclusive and six exclusive
signal regions have been studied in the ATLAS search, all defined by a
multitude of selection cuts (some more details on these issues have been given in
section~\ref{sec:monojet}).

In order to study whether any further sensitivity
can be gained, we performed a multivariate analysis (MVA) relying on a
boosted decision tree (BDT). We made use of the TMVA
framework~\cite{2007physics...3039H} and considered 14 kinematic variables,
namely the missing transverse energy $\slashed{E}_T$, the $p_T$, the pseudorapidity
and azimuthal angle of the two leading jets, the angular separation in azimuth
between the two leading jets as well as between each of them and the missing
momentum,
the angular distance in the transverse plane between the two leading jets,
the transverse mass reconstructed from each of the two leading jets and the
missing transverse momentum, the angular separation in azimuth
between the two leading jets as well as between each of them and the missing
momentum,
the angular distance in the transverse plane between the two leading jets,
the transverse mass reconstructed from each of the two leading jets and the
missing transverse momentum and the aplanarity \cite{Chen:2011ah}.

Throughout our MVA analysis, we carefully treated the issue of the
overtraining and we internally performed the Kolmogorov-Smirnov (KS) test.
Whereas in principle, the KS probability associated with both the signal and the
background needs to lie between 0.1 and 0.9, a critical KS probability larger
than 0.01~\cite{KS} and not exhibiting oscillations is also acceptable.

For our BDT analysis, we have only considered the two dominant backgrounds
consisting of invisible $Z$-boson plus jets and leptonic $W$-boson plus jets events,
carefully taking into account their individual weights. For our first signal
sample, we imposed a very loose selection cut before using the BDT,
constraining the $p_T$ of the leading jet to be larger than 150~GeV and the
number of jets to be at least 2. Choosing a benchmark setup with
$\Lambda_3 = 700$~GeV, $m_A=780$~GeV and $m_{\chi}=190$~GeV, we obtain a
number of signal and background events of $N_S = 645$ and $N_B = 9124$
respectively for an integrated luminosity of 3.2~fb$^{-1}$. The BDT analysis
yields a significance $\sigma_0 = N_S/\sqrt{N_S + N_B} = 6.49$ for a ratio
$N_S/N_B$ of about 7\%, assuming zero systematic uncertainties.
Upon considering a flat 10\% systematic uncertainty on the SM backgrounds, the
significance drops to $\sigma_{10} = 0.70$, which is almost exactly
the same as the one obtained from the best signal region in the cut-based
analysis of ATLAS. As a final check, we relaxed the initial $p_T$
selection cut on the first jet to 80~GeV and observed that the $N_S/N_B$ ratio
drops to about 3\% and $\sigma_0$ to $6.17$.

These findings lead us to the conclusion that the ATLAS cut-based analysis is already highly optimised and, as our
multivariate analysis has shown, further improving upon it appears to be highly non-trivial.

\clearpage
%=========================================================================================
\bibliographystyle{JHEP}
\bibliography{biblio}

\end{document}